\documentclass[11pt]{elsarticle}
\usepackage{extsizes}
\usepackage{graphicx}% Include figure files
\usepackage{dcolumn}% Align table columns on decimal point
\usepackage{bm}% bold math
\usepackage[mathlines]{lineno}% Enable numbering of text and display math
\usepackage{amsmath,amsthm,amssymb}
\usepackage{cuted}
\usepackage{float}
\usepackage{hyperref}
\usepackage{makeidx}
\usepackage{layout}
\usepackage{amsfonts}
\usepackage[T1]{fontenc}
\usepackage{fullpage}
\usepackage{tikz}
\usetikzlibrary{shapes,arrows}

\newcommand{\retab}[1]{[\ref{#1}]}

\newcommand{\vect}[1]{\ensuremath{\mathbf{#1}}}
\newcommand{\uni}[1]{\ensuremath{\mathbf{\hat{#1}}}}

\newcommand{\bea}{\begin{eqnarray}}
\newcommand{\eea}{\end{eqnarray}}

\newcommand{\derp}[2]{\ensuremath{\frac{\partial #1}{\partial #2}}}

\newcommand{\ori}{\ensuremath{\uni{u}}}
\newcommand{\orik}{\ensuremath{\uni{u}_k}}
\newcommand{\orii}{\ensuremath{\uni{u}_i}}
\newcommand{\vso}{\ensuremath{V_{so}(\vect r_{k,j}, \vect r_{\neig(k),k})}}

\newcommand{\orij}{\ensuremath{\uni{u}_j}}

\newcommand{\starx}{\ensuremath{[ \nskk \cdot \vect r_{kj}]}}
\newcommand{\nk}{\ensuremath{\vect n_k}}
\newcommand{\nkj}{\ensuremath{\vect n_{k,j,k}}}
\newcommand{\nskk}{\ensuremath{\vect n_{\neig(k),k,k}}}
\newcommand{\nj}{\ensuremath{\vect n_j}}
\newcommand{\vrkj}{\ensuremath{\vect r_{k,j}}}

\newcommand{\neig}{\ensuremath{s}}
\newcommand{\vrnkk}{\ensuremath{\vect r_{\neig(k),k}}}

\newcommand{\beq}{\begin{equation}}
\newcommand{\eeq}{\end{equation}}
\makeindex

\title{A computational model relating the self-assembly in a fluid of lath like particles with its rheology and gelation}

\author{G. Villalobos}
\ead{gabriel.villalobosc@utadeo.edu.co}
\address{Universidad de Bogot\'a Jorge Tadeo Lozano, Departamento de Ciencias B{\'a}sicas. Calle 22 N{\'u}mero 10 - 30, Edificio Manrique. 110311. Bogot{\'a}, Colombia.}
\address{Computational Biophysics, University of Twente, P.O. Box 217, 7500 AE, Enschede, The Netherlands}

\begin{document}

%\linenumbers\relax % Commence numbering lines
\begin{abstract}
We study the self-assembly leading to a gel transition occurring in a numerical model of a solution of slender, colloidal sized particles, called laths, who interact mostly in the direction perpendicular to their areas. At the particle level, the attraction causes them to align into long aggregates of several particles, called whiskers in the literature.  To simulate the process, we have developed a Brownian dynamics model in which the attractive interaction comes from a potential energy that depends on both the relative orientation of the laths as well as normal vectors to their areas, disregarding their width. The simplicity of the model allows the simulation to reach large enough times, of the order of minutes, needed to simulate numerical rheology tests. With this we are able to characterize the whisker formation, as well as to simulate the gel transition.  A a conclusion of this work, we have shown that the gel transition can occur even if the whiskers are not allowed to branch, as is the case in this model.
\end{abstract}


\date{\today}

\maketitle
%\nolinenumbers
\tableofcontents
%\linenumbers\relax % Commence numbering lines

\section{Introduction}
In the present letter, we want to study both the structure and
rheology of aggregates of long slender particles using a computational
model. This aggregates are called whiskers in the literature. In order
to retain just the basic elements of the system, the particles are
simulated as a slender plane, a lath, that is defined by just two
directions, one along the long axis and one perpendicular to it. The
potential energy that generates the stacking depends on only one
energetic parameter. It consists on the product of two functions, one
that depends on the relative orientation among laths and one that
represents an excluded volume contribution. In combination with a
Brownian dynamics method, this allows to produce a fast and simple
model which generates and allows to study the aggregation of the laths
into whiskers. Our lath model deviates from the Brownian dynamics simulations that study the aggregation of rod like particles, as in \cite{C5SM01845E}. 
\begin{figure}[htbp]
  \centering
  \includegraphics[width=7cm,bb=0 0 585 441]{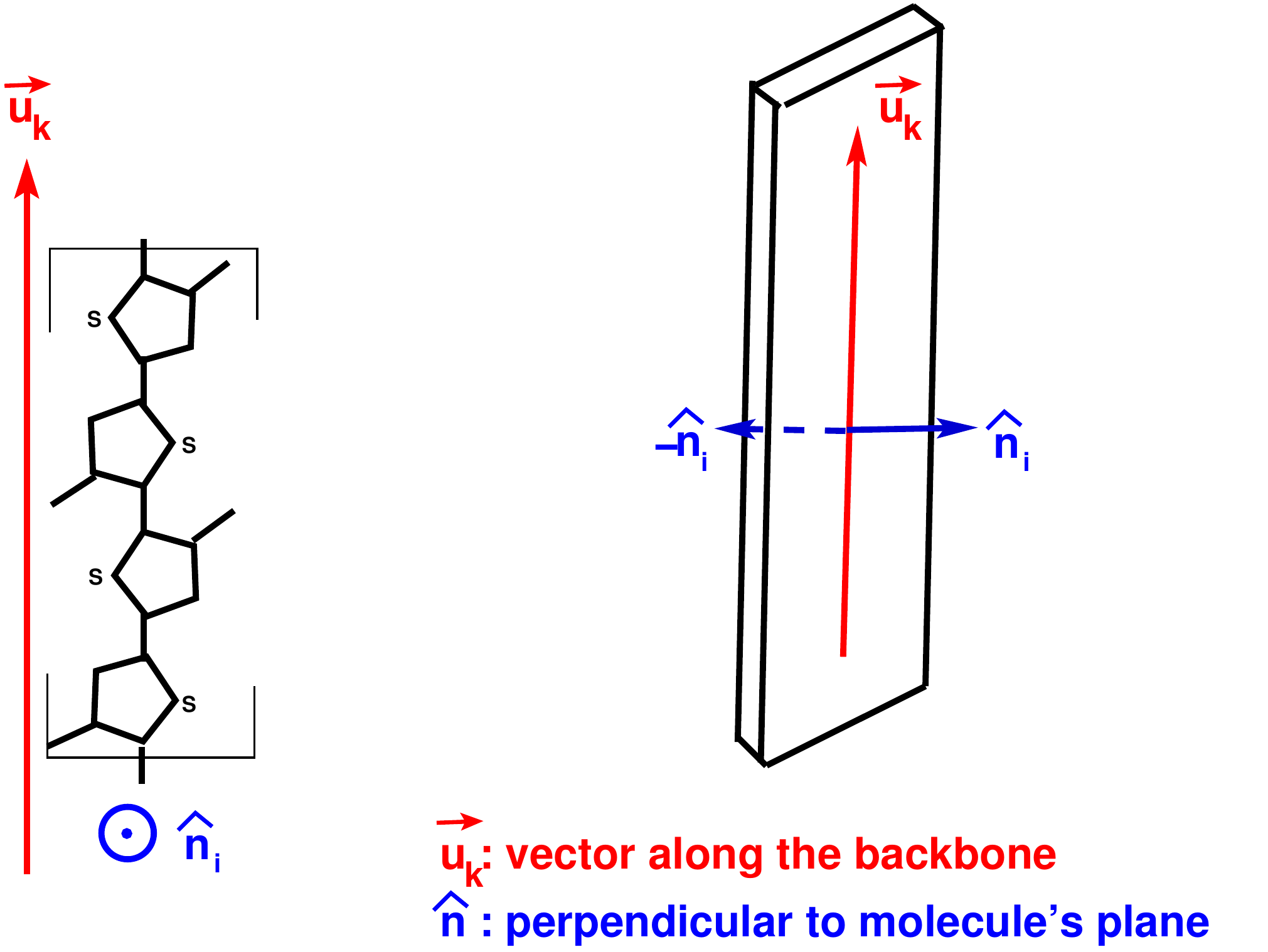}
  \caption{
    \label{modelMolecule.pdf} % FIG1
    Model of the molecule.  The direction $\orik$, in red, is
    defined along the backbone of the P3HT molecule.  The normal
    $\uni n$, in blue, is defined perpendicular to the plane of the
    molecule. \textit{{(Left)}} The backbone is specifically
    depicted. \textit{(Right)} In the three dimensional
    representations there are two $\uni n$ vectors. They are further
    explained in \ref{threebodypot.pdf}
  }
\end{figure}

In order to provide an experimental system, we will make reference to
the aggregates of P3HT molecules. For that particular case our model
is just a rough estimate, since we will not be taking into account the
flexibility of such molecule. Furthermore the whole molecule should be
coarse grained into a single lath, which can be seen as a very bold
approximation. However, our numerical simulations show a very good
resemblance to the rheology of such system, which shows that the
highly coarse graining is not far from the physics of this phenomenon.

The strong forces between molecules (or macromolecules) who have a
$\pi$ structure is known in the literature as \emph{aromatic}, or
simply $\pi-\pi$ interaction \cite{doi:10.1021/ja00170a016}. In the
case of some of semi-flexible polymers, it is known to cause stacking
of the aromatic groups; which in turn drives a self-assembly process
that creates structures orders of magnitude larger than the length of
the original particles\cite{Lim201014,POLB:POLB22310}.

An important process that develops $\pi$ stacking is the polymer
component used in creation of hetero-junctions of organic solar cells
being, for instance, a mixture of P3HT and $C_{60}$
\cite{Nelson2011462,sirringhaus1999twodimensionalcharge}. One
particularly interesting problem, both from the simulation and the
experimental point of view, is to understand the process of formation
of the interface between the two components; as it affects the
performance of the capture of the energy by the solar cell
\cite{Geiser2008464}. For instance, some experimental groups have
studied the relationship between the molecule characteristics with the
final behavior of the cell is of special relevance \cite{Li2013}. \\
\begin{figure}[htbp]
\centering
\includegraphics[width=16cm]{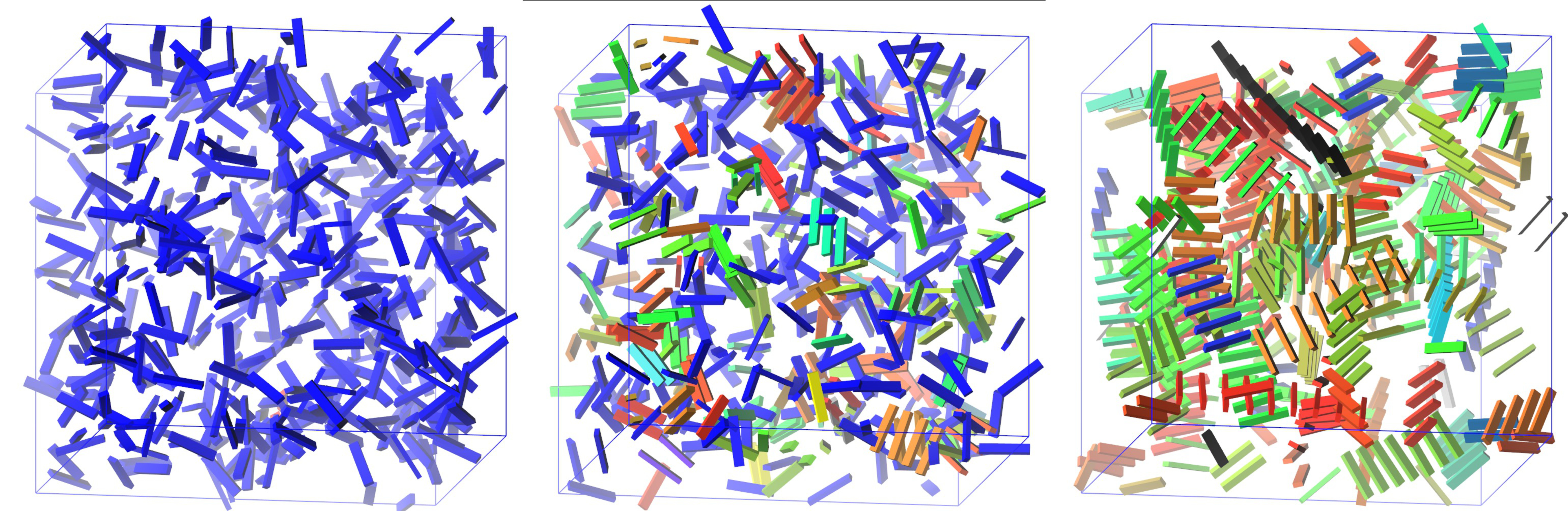} %FIG2
\caption{
\label{495_T_40_three_plots.png}
Snapshots of one
  simulation. T = $40$ $K$, and the rest of the
  parameters are those of \retab{table:parameters}. \textit{(Left)}:
  Initial state of the simulation.  The color code corresponds to
  different whiskers, with blue being the color of free
  laths. \textit{(Middle)}: $t = 0.5$ $s$, starting to create some
  whiskers.  \textit{(Right)} : At $t = 200$ $s$, several long
  whiskers.  (\textit{Color online.})}
\end{figure}

On the macroscopic level, the self-assembly process generates a change
in the rheology: the gelation of the solution
\cite{doi:10.1021/ma9005445}. One particularly clever experimental
set-up consists in probing this network by comparing measurements of
the electrical conductivity and the rheology of such gels
\cite{doi:10.1021/ma202564k,doi:10.1021/ma2000515}. We are able to
pursue similar rheological tests with our present computational
model. Specifically, we can test the loss and storage modulus of the
system while changing the temperature. This provides a way to test the
possibility of gelation.

Previous computational studies (Langevin coarse grained and MD
simulations) have studied both the self-assembly process of the P3HT
molecules into whiskers,
\cite{C3NR33324H,lee2012ellipsoid,doi:10.1021/jp907338j}; as well as
the gel formation \cite{doi:10.1021/ma302343e}. Although successful,
these models have the drawback of relying in complex force fields;
which are both computationally expensive and difficult to interpret by
isolating the individual interactions. To the best of our knowledge,
no similar coarse grained simulations of stacking of particles have
been attempted, either in polymer or colloidal systems.

\section{Model and simulations}
We propose the model to be named Brownian Orientational Lath Model,
and to use the acronym BOLD. 

\subsection{Configurations and interactions}
As already emphasized above, our goal is to investigate the gel
transition of strongly interacting lath-like objects of colloidal
sizes. Although the basic system to which we apply our model will be
that of $\pi$-stacking polythiophenes in solution, we will keep the model
itself as generic as possible. We therefore consider as basic object
of our simulations a lath as depicted in
\ref{modelMolecule.pdf}. We describe its position the three
Cartesian coordinates $\vect r=\left\{x, y, z\right\}$ of its center
of mass, and its orientation in space by and the two perpendicular
unit vectors $\uni u$ and $\uni n$. Subscripts will indicate the
particular lath under consideration. The vector $\uni u$ specifies the
direction of the long axis of the lath while $\uni n$ is chosen to be
perpendicular to the plane spanned by the two longest edges of the
lath. In all that follows it is assumed that the shortest of the three
edges is much shorter than the others, as is the case in P3HT.

We propose to describe the potential of a given configuration as  
\beq
\label{eq:totalALIGpot} \Phi_{S} = \epsilon
\sum_{j,k} V_{d}(r_{kj})V_{o}(\orik,\orij) V_{cn}(\uni n_k \cdot
\vect r_{kj})V_{cn}(\uni n_j \cdot \vect r_{kj}), \eeq 

\noindent where $\vect r_{kj}=\vect r_k - \vect r_j$, is connector
between two laths, and $r_{kj}$ is its length. The first factor in
each term describes the dependence of the interaction energy on the
distance between the centers of mass of the two laths under
consideration, and the three other factors describe the dependence on
their relative orientations. The explicit expressions are as follows
\bea \nonumber V_d(r_{kj}) &=& %\begin{cases}
\frac{1}{2}
\left(\frac{\tanh \left(a\left(r_{kj} -\frac{\sigma}{2} \right)\right)}{\tanh \left(\frac{a \sigma}{2} \right)} -1 + \frac{A_{excl}}{(r-r_{excl})^4}\right)\\ 
\nonumber
     V_{o}(\orik,\orij) &=& (\orik \cdot
     \orij)^{2}\\ \nonumber V_{cn}(\uni n_{k}, \uni r_{kj})
     &=& \begin{cases} (\uni n_{k} \cdot \uni r_{kj})^{2} & \mbox{
         if } \uni n_{k} \cdot \uni r_{kj} > 0 \\ 0 &\mbox{ else
     } \end{cases}.  \eea

\begin{figure}[htbp]
\centering
\includegraphics[width=7cm]{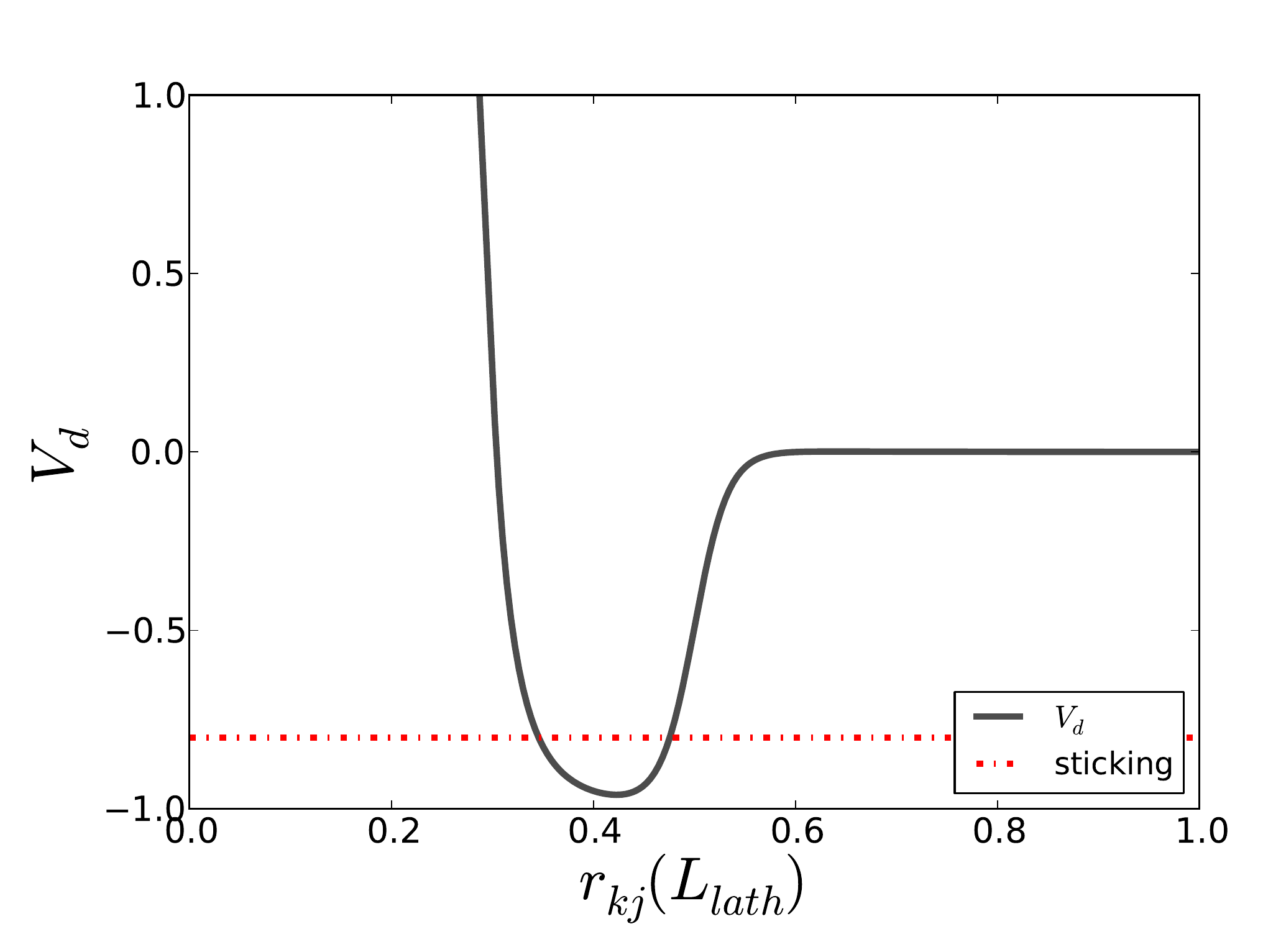} %FIG3
\caption{
  \label{Vdist.pdf}
  The dependence of the potential with the
  distance,$V_d(r_{k,j})$, is a narrow function with its center close
  to $0.4$ $L_{lath}$. Below a certain cutoff, $r_{cutoff}=0.3$  $L_{lath}$, it is replaced
  with a linear extension. The threshold for sticking and
  unsticking is close to the bottom of the well, that is $0.2
  \epsilon$ above the absolute bottom of the well, in the dash-dotted red
  line.  (\textit{Color online.}) }
\end{figure}

The second factor, $V_{o}(\orik,\orij)$, favors a parallel orientation
for the long axis of the two laths. The last two factors, $V_{cn}(\uni
n_k \cdot \vect r_{kj})$ and $V_{cn}(\uni n_j \cdot \vect r_{kj})$,
minimize the energy when the two vectors $\uni n_k$ and $\uni n_j$ are
both parallel to the connector $\vect r_{kj}$, thereby favoring the
two faces of the laths to be parallel. The reason for distinguishing
between positive and negative values of $\uni n_j \cdot \vect r_{kj}$
will be explained below. Expression $V_d(r_{kj})$ is plotted in
\ref{Vdist.pdf}. The minimum is seen to be at a distance
between two laths of about $0.4*L_{lath}$. This implies that the
thickness of the lath plus the face-to-face distance is about equal to
this value. In the simulations, the potential is linearized
below a $0.3*L_{lath}$ to allow for acceptable time steps. The values
of the various parameters in $V_d(r_{kj})$ are given in
\retab{table:parameters}.

Within a whisker, a lath is strongly bound to one; or, more commonly,
to two other laths. In order to complete the description of the
potential energy, we must define when this binding within two laths
happens. This will be the case when their interaction energy is less
than a threshold energy $V_{thr}$ for which we choose $V_{thr}=-0.8
\epsilon$. Then, two laths which are strongly bound to each other,
while none of the two laths is strongly bound to another lath,
interact like a pair of free laths.

\subsection{Propagator and refinement of potential} 

In general there is no qualitative difference between the dynamics of
$\orik$ and that of $\uni n_k$, a simulation shall treat both with the
same methods and precision. Notice, however, that $\uni n_k$ is by
definition perpendicular to $\orik$, so that $\orik$ has three degrees
of freedom and $\uni n_k$ has one less. In the case of lath-like
particles, rotations around the long axes will be much faster than
reorientations of $\orik$ or displacements of the laths as a whole; in
other words, the dynamics of $\uni n_k$ will be much faster than that
of $\orik$. We will therefore assume that during one time step the
lath has fully explored all possible orientations around its long axis
and has settled into the corresponding energetic minimum. This means
that during the simulation the vector $\uni n_k$ will be determined by
the configuration described by $\left\{ \vect r_k, \orik \right\}$ and
its recent history (see below). We are then left with two equations of
motion for each particle: \bea
\label{eq:equmotion}
  d\mathbf{r}_k
    & = & - \frac{1}{\xi_0}
             \nabla_k \Phi_C dt +  \nabla_k \left( \frac{k_B T}{\xi_0} \right) dt
      +   \pmb{\Theta}_k^t \sqrt{ \frac{2 k_B T dt}{\xi_0} }.  \\
 d\ori_k & = & 
\frac{L_{lath}^2}{9 \xi_0} \vect T \times \ori_k dt + \pmb \Theta_k^r\frac {L_{lath}}{3}\sqrt{ \frac{2 k_B T dt}{\xi_0} }
\eea

Here $\pmb{\Theta}_k^t$ is a three dimensional random vector with
components which have zero mean and unit variance, uncorrelated among
each other. Similarly $\pmb{\Theta}_k^r$ is a two dimensional
random vector with uncorrelated components having zero mean and unit
variance. The two random rotations are applied around two
perpendicular axes orthogonal to $\ori_k$. After each time step the
length of the vector $\ori_k$ is brought back to unity by a shortening
along the original orientation \cite{COBB05,so55984}. $\xi_0$ is the
average translational friction of the lath; we have chosen the
rotational friction to be related to the translational friction as for
infinitely long rods \cite{Doi,so55984}.

Let us now describe how we update $\uni n_k$. First, consider two
laths $k$ and $j$, neither of whom is interacting with any other
lath. We choose $\vect n_k=\vect r_{kj}-(\vect r_{kj}\cdot
\ori_k)\ori_k$, and similarly for $\uni n_j$, since these are the
orientations assumed to correspond to the minimum interaction between
the two laths for the given configuration $\left\{ \vect r_{kj},
\ori_k, \ori_j \right\}$; unit vector $\uni n_k$ is $\vect n_k /
\sqrt{\vect n_k\cdot \vect n_k}$. This choice will drive the two laths
to become parallel and have their centers of mass connector $\vect
r_{kj}$ perpendicular to both $\uni u_k$ and $\uni u_j$. Substituting
these expressions for $\uni n_k$ and $\uni n_j$ into the potential
above leaves us with a simple pair potential as commonly used in
simulations of hard convex bodies.

The situation is different when a free lath $j$ interacts with a lath
$k$, which is strongly interacting with another lath $\neig(k) \neq
j$. In this case lath $j$ may still freely reorient along its long
axis, but lath $k$ is restricted to do so by its strong interaction
with $\neig(k) \neq j$. We therefore choose $\uni n_j$ as above, but
set $\uni n_k$ equal to the normalized $\nk = \vrnkk - (\vrnkk \cdot
\uni u_k) \uni u_k$. The $\neig(k)$ is strongly bound to lath $k$. We
are now in the situation where we need to distinguish between two
cases in the definition of $V_{cn}$. Only when lath $j$ is situated on
the positive side of lath $k$ it may bind to lath $k$; on the negative
side lath $k$ is already bound to lath $\neig(k)$ and cannot bind to
the newly arriving lath $j$ anymore. Notice that this treatment of the
re-orientations of the laths around their long axes turns the
potential in this case into a three body potential. Moreover the
potential becomes history dependent. This is similar to coarse grain
simulations of linear polymers \cite{0953-8984-23-23-233101}, where
the entangling of polymers is not fully determined by the
configuration of the beads, but depends on how the beads came to that
particular configuration.

\begin{figure}[htbp]
\centering
\includegraphics[width=7cm,bb=0 0 899 395]{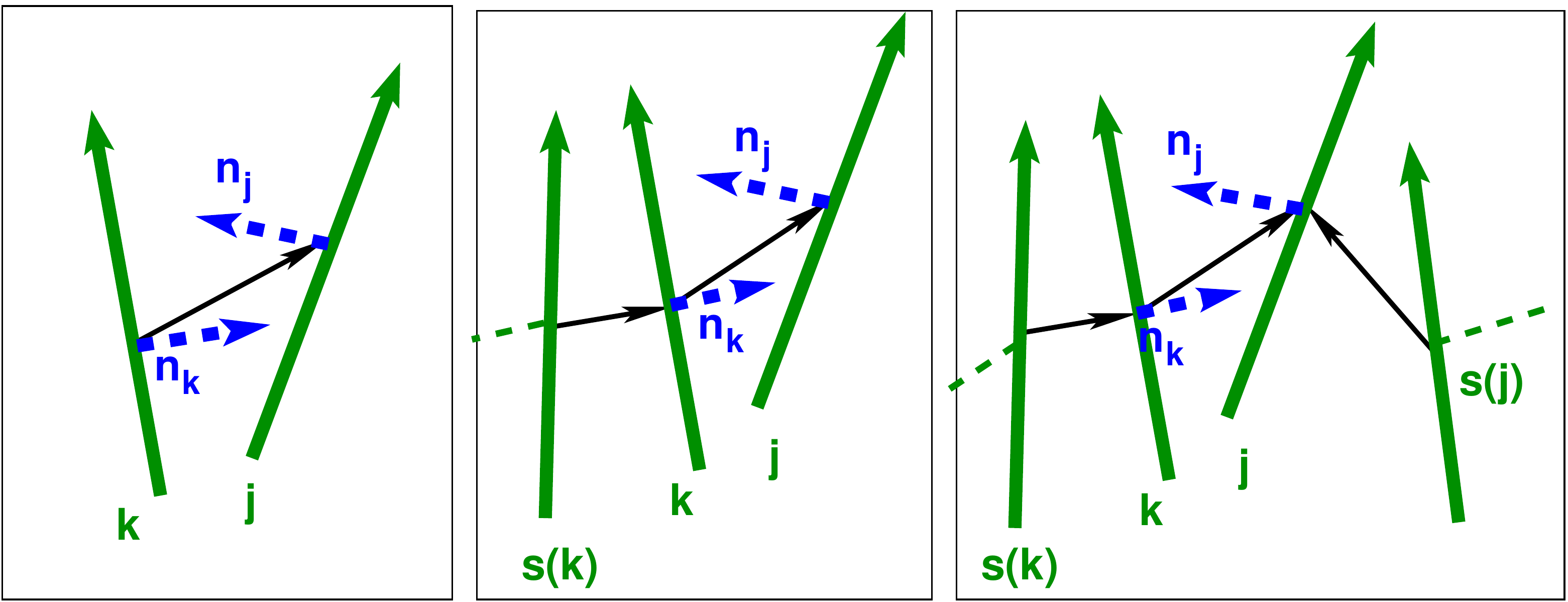} %FIG4
\caption{
\label{threebodypot.pdf}
Sketch of the vectors used in the
  three body potential. The notation is: in bold continuous (green),
  the orientational vectors $\orik$; in thin continuous (black), the
  connectors; in thick dashed (blue), the normal vector $\nk$; finally
  in thin dashed (green), possible connectors to other laths.  The
  mathematical model in this letter is three dimensional but, to
  simplify the sketch, all vectors of this plot lie in the same plane.
  $\neig(k)$ is stuck to $k$, and $k$ and $j$ may or may not be stuck
  together. (\textit{left}) Neither $k$ nor $j$ is stuck to any other
  lath (say $l$), then we pick $\nk$ and $\nj$ using the components of
  the connector that are perpendicular to the respective orientation
  vectors. (\textit{center}) $k$ is stuck to $\neig(k) \neq j$, $j$
  does not have another stuck particle; the normal in $k$ is given by
  its interaction with its stuck particle $s(k)$. (\textit{right})
  Both $k$ and $j$ are stuck to some other laths. (\textit{Color
    online.})}
\end{figure}

A third situation occurs when two laths $k$ and $j$ approach each
other, but are already bound to $\neig(k) \neq j$ and $\neig(j) \neq
k$ respectively. In this case we have $\nk = \vrnkk - (\vrnkk \cdot
\uni u_k) \uni u_k$ and similarly for $\nj$. Only when lath $j$ is on
the positive side of lath $k$, determined by lath $\neig(k)$, and lath
$k$ is on the positive side of lath $j$, determined by $\neig(j)$,
will the interaction be non-zero. This treatment turns the interaction
into a four-body interaction. The description given above is summarized in \ref{threebodypot.pdf}. 

\subsection{Parameter settings}
A set of nondimensiolanized units are given by the average
translational diffusion coefficient of the laths $D_0=k_BT_0/\xi_0$,
the thermal energy $k_BT_0$ and the length of a lath $L_{Lath}$. With
this choice, the unit of time is $L_{Lath}^2/D_0$, the unit of
velocity is $D_0/L_{Lath}$, and the unit of pressure is
$k_BT/L_{Lath}^3$. However, SI units are used in the description of
the results of this paper.

All model parameters used in this study are given in
\retab{table:parameters}, once in reduced units and once in SI
units. The latter are only used in a few occasions to show that with
reasonable parameters the model gives results that are comparable to
those found experimentally. The SI parameter settings are meant to be
reasonably close to those that apply to P3HT solutions, even though
the model that we present here can be used in other contexts. 
We emphasize that the important characteristics of the model are that
bonding can only occur in one-dimensional structures and that the
bonding is very strong, $i.e.$ $\epsilon=100k_BT_0$.

The length of the molecule was chosen to match commercially available
P3HT that was reported to us in private communication from an
experimental group at TU/e.
\begin{table}[h]\footnotesize \caption{Parameters used in the model}\label{table:parameters}
  \begin{center}
    \resizebox{7cm}{!}{
      \begin{tabular}{| p{3.2cm}|  c|  p{3.2cm}| }
        \hline
        Parameter& Symbol & Value \\ \hline\hline
        Strength of the alignment potential.  & $\epsilon$ & $4.04 \times 10^{-19} J$ $=100 k_B \times T_0$\\ \hline
        Length of the laths, unit of distance &$L_{lath}$ & $68.44$ $nm$ $\approx 177$ thiophene rings \\ \hline
%        & & \\ \hline
        Length of the size of the squared shaped simulation box& $L_{box}$ &  $ 8\times L_{lath} = 0.5452$ $\mu m$\\ \hline
        Laths in the simulation box & $N_{laths,total}$ &   $512$ \\ \hline
        Inflection point of $V_d$ & $\sigma/2$ &   $34.22$ $nm$\\ \hline
        Cutoff for the linear extension of $V_d$ &  $r_{cutoff}$  & $0.3  L_{lath}\approx 20.53 nm$ \\ \hline
        Threshold for sticking/unsticking of laths. &  $V_{thr.}$ & $-0.8\epsilon$ \\ \hline
        Slope (sharpness) of potential $V_d$ & $a$ & $ 20 m^{-1}$ \\ \hline
        Excluded volume distance constant &  $r_{excluded}$  & $0.22 L_{lath}\approx 15.05 nm$ \\ \hline
        Excluded volume amplitude &  $A_{excluded}$  & $0.0001$ \\ \hline
        Power of the orientational potential  & $l$ & $1$ \\ \hline
        Time-step & $\Delta_t$ & $1 \times 10^{-5} s$ \\ \hline
        Initial Temperature & $T_0$ & $293$ $K$ \\ \hline
        Solvent friction (translational) & $\xi_0$ & $8.66 \times 10^{-7} kg / s$ \\\hline % (2.8368e-3+1)*(SI_ti*BOLTZ*SI_te)/SI_r**2.
        Solvent friction (rotational) & $\xi_{0t} = \frac{\xi_0}{9}$ & $9.62 \times 10^{-8} kg / s$ \\\hline % (2.8368e-3+1)*(SI_ti*BOLTZ*SI_te)/SI_r**2.
    \end{tabular}}
  \end{center}
\end{table}

Results from a typical run with these parameters and at temperature
$T=40$ $K$ are shown in \ref{495_T_40_three_plots.png}. In the
left panel a box is shown at the beginning of the run. All laths are
colored blue indicating that none of them is strongly bound to any of
the others. In the middle panel the same box is shown at time $t=
50000 $, which corresponds to $0.5$ $seconds$ in SI units. It is
clearly seen that several laths have collected into short chains, also
called whiskers from now on. Laths which are connected through a
sequence of consecutive strong bonds are given the same
color. Different colors correspond to different whiskers. In the right
panel, the same box is shown at $t= 20000000$, which corresponds to
$200$ $seconds$ in SI units. The box has now reached equilibrium and
several long whiskers are observed.

\section{Results}

In this section we describe the results of our simulations. First we
describe the approach to equilibrium of simulations boxes which were
initially either fully disordered or fully ordered. This gives
information about typical time scales in the system. We also briefly
analyze the structure of final equilibrium boxes. In a second
subsection we describe the rheological properties of our systems, with
special attention being given to the gel transition.

\subsection{Time evolution of formation of whiskers and equilibrium distributions of lengths}

As we have seen in the previous section, the model parameters chosen
in this study lead to the formation of whiskers during simulations
starting at time zero with completely disordered boxes. In this
section we analyze how this structure evolves with time. To fully
characterize the structure of the whiskers we studied several
characteristics: its length as a function of time,in number of laths
per whisker; the number of whiskers formed in the simulation box, as a
function of time; the total material in whiskers as function of time,
and the histogram of distribution of whisker sizes. 

\begin{figure}[htbp]
\centering
% pdfjam 848_847_846_845_844_843_bigRUNS__meanNwhisker_several.pdf 842_841_840_839_838_837_bigRUNS__meanNwhisker_several.pdf --nup  2x1 --landscape --outfile Page1+2.pdf
\includegraphics[width=14cm,bb=0 0 842 595]{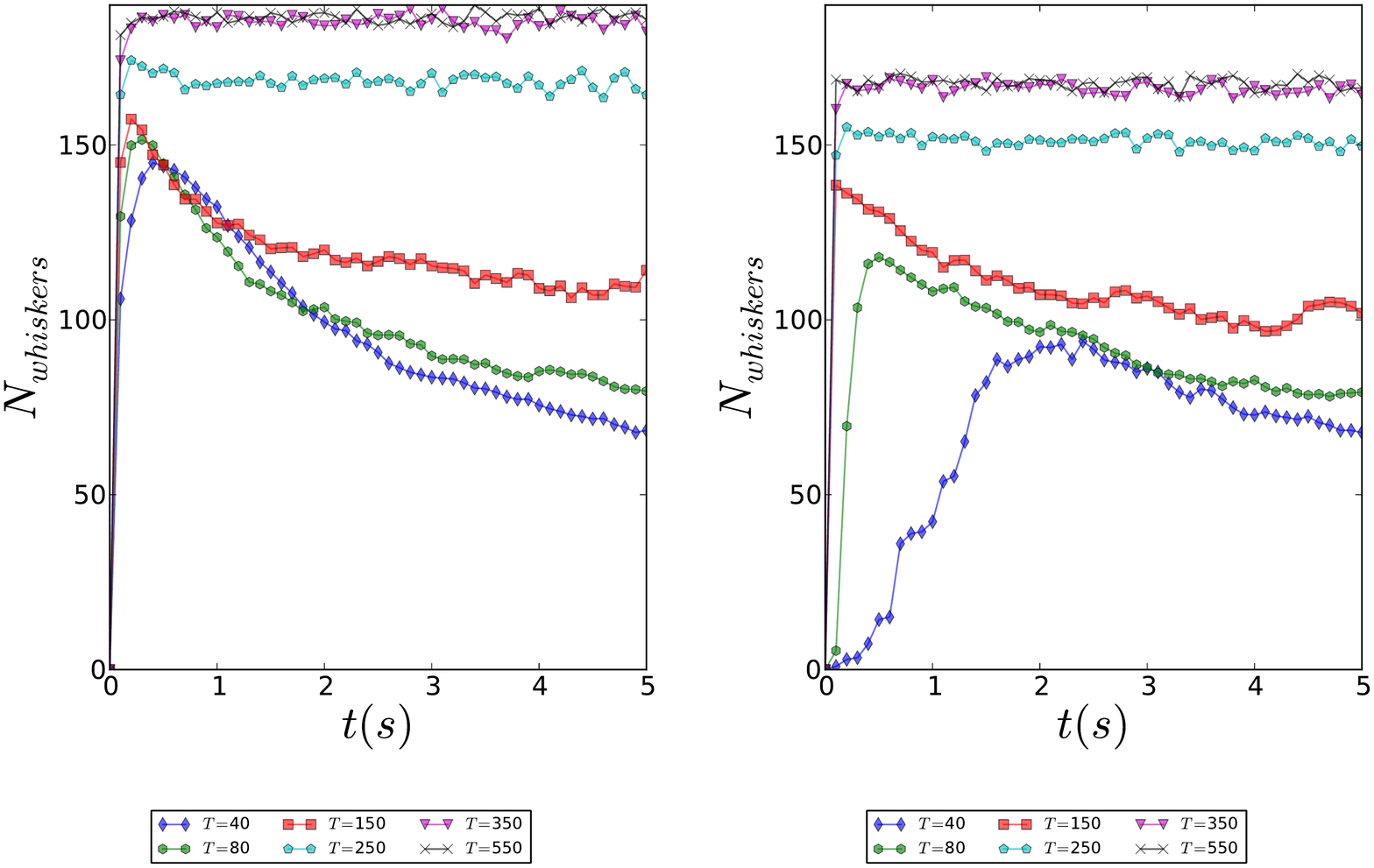} %FIG5
\caption{
\label{848_847_846_845_844_843_bigRUNS__meanNwhisker_several.pdf}
  Number of whiskers as function of time, for 6 different
  temperatures (in Kelvin); first $5$ $s$; average over 20 simulations.  \textit{(Left)}
  initially disordered configuration, the number of whiskers reaches a
  higher plateau for higher temperatures. \textit{(Right)} initially ordered
  configuration. The only discernible difference between ordered and
  disordered initial configurations after a few seconds is a lower
  plateau for the high temperature average number of whiskers in the
  case of initially ordered simulation.  (\textit{Color online.})}
\end{figure}

In
\ref{848_847_846_845_844_843_bigRUNS__meanNwhisker_several.pdf}
we present, for various temperatures, the time evolution of the number
of whiskers in our boxes. The left panel corresponds to those that
started with randomly distributed laths, while the right hand side
panel presents data for boxes starting from fully ordered boxes,
\textit{i.e.} boxes in which all laths were collected in one long
whisker. Only the first four or five seconds are shown. The initially ordered simulations serve as a test of the consistency of the model. 

For temperatures substantially larger than $T_0$, large numbers of
whiskers (necessarily small) are observed after the first one tenth of a
second. In boxes starting from disorder these are obtained by
aggregation of laths, while in boxes starting from the ordered
structure these are obtained by disintegration of the initially
available very long whisker. After the first one tenth of a second
hardly any changes occur in the number of whiskers in these
boxes. Notice that the various plateau values are not equal for
corresponding temperatures in the left and right hand side
panels. This is due to the fact that with our code the identification
of whiskers is history dependent, as already mentioned in the previous
section. This somewhat unrealistic aspect only affects the numbers of
very short whiskers.

For temperatures of the order of $T_0$ or less, in boxes starting from
disorder, again a large number of whiskers is formed within the first
few tenths of a second. From then on the number of whiskers start to
gradually decrease with increasing time. This is due to the fact that
initially only very small whiskers are formed, which then start to
merge into longer whiskers. Eventually, equilibrium distributions of
lengths occur, depending on the temperature. For these temperatures,
boxes starting from ordered configurations behave only slightly differently
than those starting from disorder. Again, as in the high temperature
case, the long initial whiskers start to disintegrate, but this time
increasingly slower with decreasing temperatures. This difference in speed of disintegration is not very important, as we do care most of the equilibrium distribution of initially disordered systems. There are still
overshoots in the numbers of whiskers after which the distributions of
whiskers begin to rearrange in order to approach the appropriate
equilibrium distributions.

In this plot, as well as the ones following, we have included a
$T=550$ $K$. The goal of making simulations at this temperature is to
show the trend that would have increasing the temperature keeping all
the other interactions unchanged. It has to be taken with caution,
though, since for a system like P3HT at this temperature one can
expect melting to happen.

\begin{figure}[htbp]
\centering
\includegraphics[width=7cm,bb=0 0 439 583]{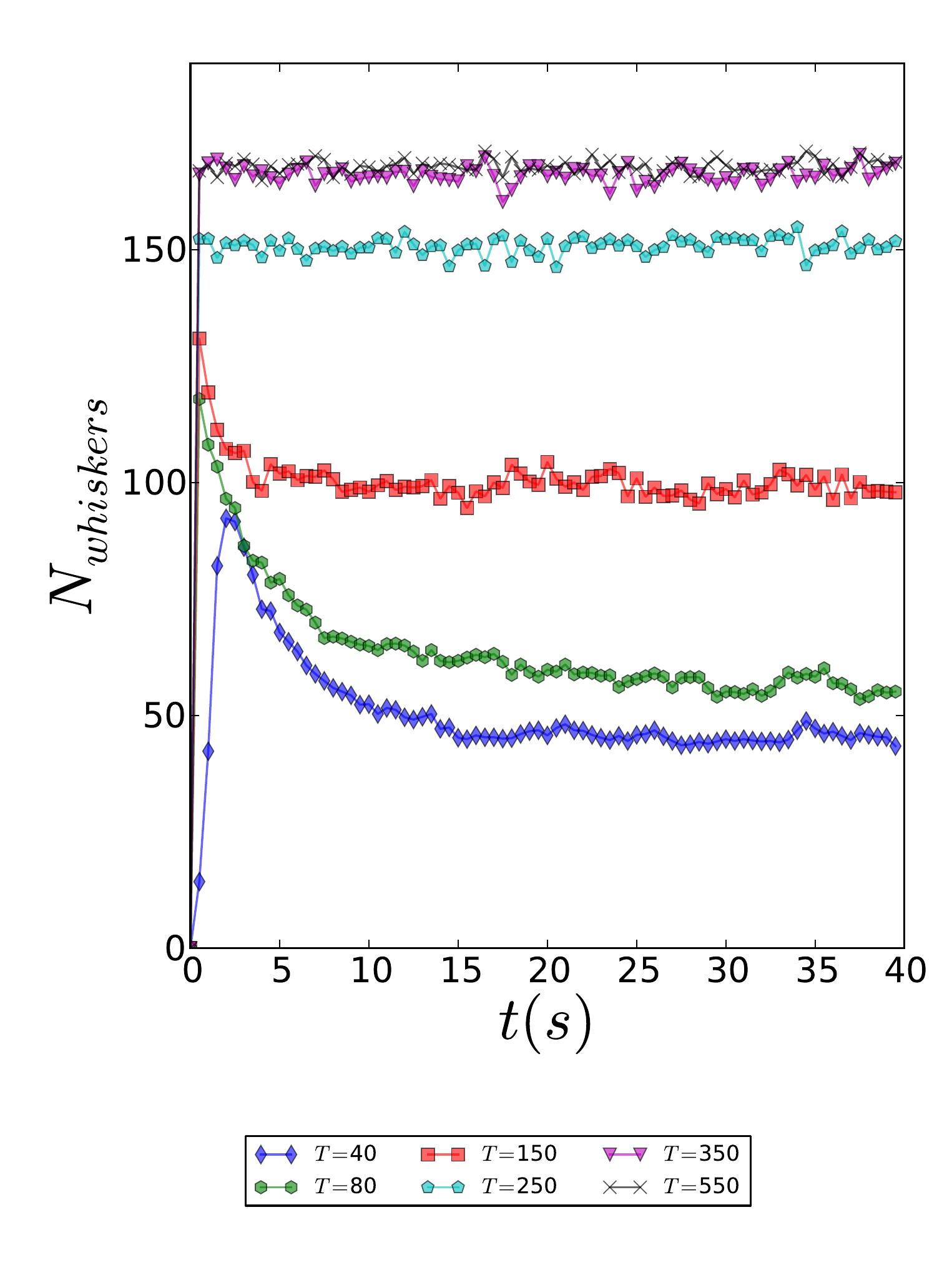} %FIG6
\caption{
  \label{842_841_840_839_838_837_bigRUNS__meanNwhisker_several_long.pdf}
Number
  of whiskers as function of time, for different temperatures; average
  over 20 simulations; initially ordered configuration. After the
  initial sharp growth in number of whiskers there is a decay to a
  plateau value that depends on the temperature.  (\textit{Color online.})}
\end{figure}

In
\ref{842_841_840_839_838_837_bigRUNS__meanNwhisker_several_long.pdf}
we again present the time evolution of the number of stickers for the
same temperatures as above, but this time for a much longer time span
of $40$ $s$. Since there is no difference after the first five seconds
between the simulations starting from disorder and those starting from
order, except for the very high temperatures, we restrict the data
here to those of boxes starting from ordered structures. The main
conclusion from theses runs is that it is safe to assume that
equilibrium distributions have developed only after about twenty to
twenty five seconds, depending on temperature.

\begin{figure}[htbp]
\centering
\includegraphics[width=14cm,bb=0 0 842 595]{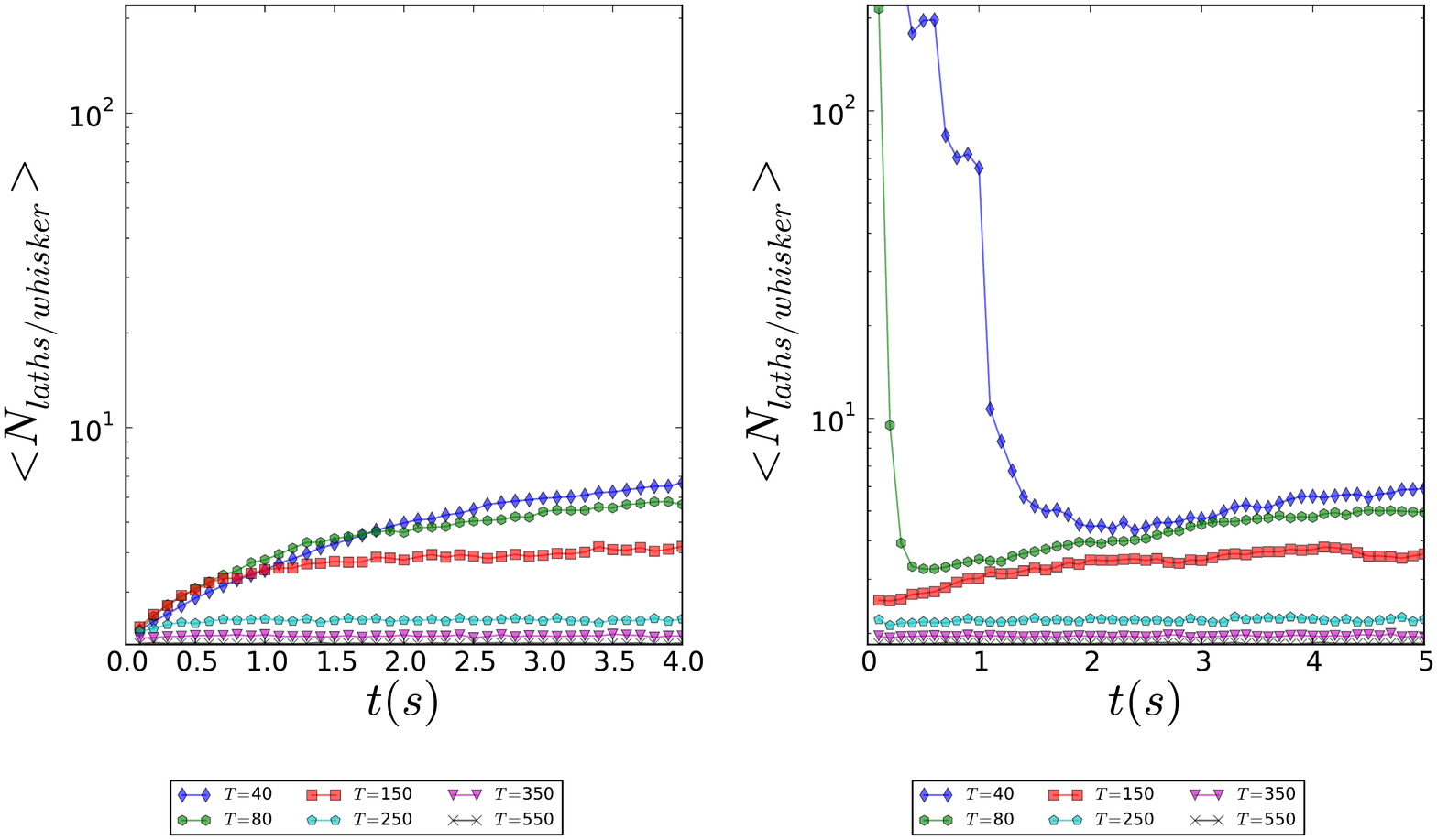} %FIG7
\caption
        {
\label{FIG7.pdf}
Average number of laths per whisker, as function of temperature;  for 20 simulations. In both panels the main axis are set in
logarithmic scale.\textit{(Left)} $4$ $s$ of the initially disordered configuration,
\textit{(Right)} $5$ $s$ of the initially ordered configuration.  (\textit{Color online.})}
\end{figure}

\begin{figure}[htbp]
\centering
\includegraphics[width=7cm,bb=0 0 441 585]{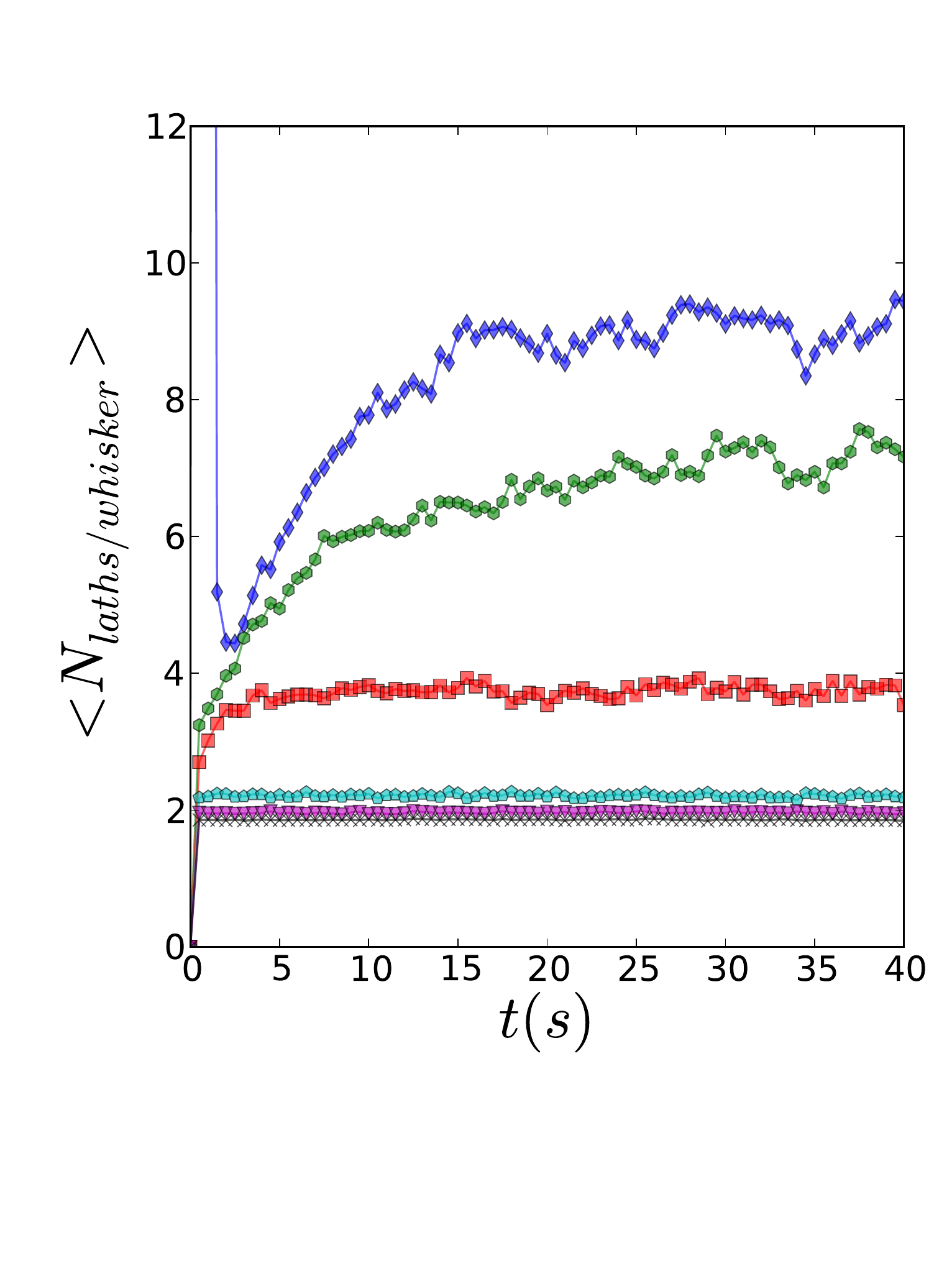} %FIG8
\caption{
  \label{842_841_840_839_838_837_bigRUNS__meanRODSinWHISKER_b.pdf}
        Average number of laths per whisker, as function of
          temperature; during $40$ $s$; initially ordered
          configuration; averaged over 20 simulations. Due to a different vertical scale, not all of the data points of \ref{FIG7.pdf} are seen here.  (\textit{Color online.})}
\end{figure}

\ref{FIG7.pdf} shows the transient behavior, initial $4$ or $5$
seconds, of the average number of laths per whisker,
$<N_{laths/whisker}>$, for both initially disordered and ordered
configurations. Notice that the vertical axes are on log scales to
more easily span a broad range of values.  Both plots show the
expected behavior: a steady increase of the number of laths per
whisker, for the initially disordered configurations; a breaking up of
the initially available whisker spanning the whole box, followed by a
steady increase of the number of laths per whisker, for the initially
ordered simulations. Again, the lower the temperature, the slower is
the time evolution of the towards the final plateau
values. \ref{842_841_840_839_838_837_bigRUNS__meanRODSinWHISKER_b.pdf}
shows the long time behavior, up to $40$ $s$, for the initially
ordered simulations. As before we notice that up to twenty five
seconds are needed before the systems reach equilibrium. Notice also
that the high temperature systems reach only average values of two
laths per whisker, the minimum value to form a whisker; as the
temperature decreases the plateau values grow to values of about ten
for the lowest temperature. As temperature increases, longer whiskers
are broken into shorter ones, thus reducing the number of laths per
whisker.

It is interesting to have a look at the time evolution of the total
number of laths taking part in whiskers. These results are shown in
\ref{FIG10.pdf}. Again, in the left panel results refer to
systems starting from disorder, while the right panel refers to
systems starting from order. It is seen that in systems starting from
disorder the total material in whiskers is monotonically increasing,
while with systems starting from ordered boxes the total material in
whiskers is almost constant; an initial large whisker may split into
some whiskers, but does not result into individual laths. It is
interesting to notice that after five seconds all simulations have
reached their steady state values. This means that after this time the
only thing that happens is a rearrangement of the laths in the
whiskers in order to approach the appropriate equilibrium
distributions. This process takes about twenty seconds. The difference of the total material in whiskers with respect to the temperature comes from the fact that at higher temperatures there is more chance that there are single laths not belonging to any cluster.

\begin{figure}[htbp]
\centering
\includegraphics[width=8cm,bb=0 0 585 441]{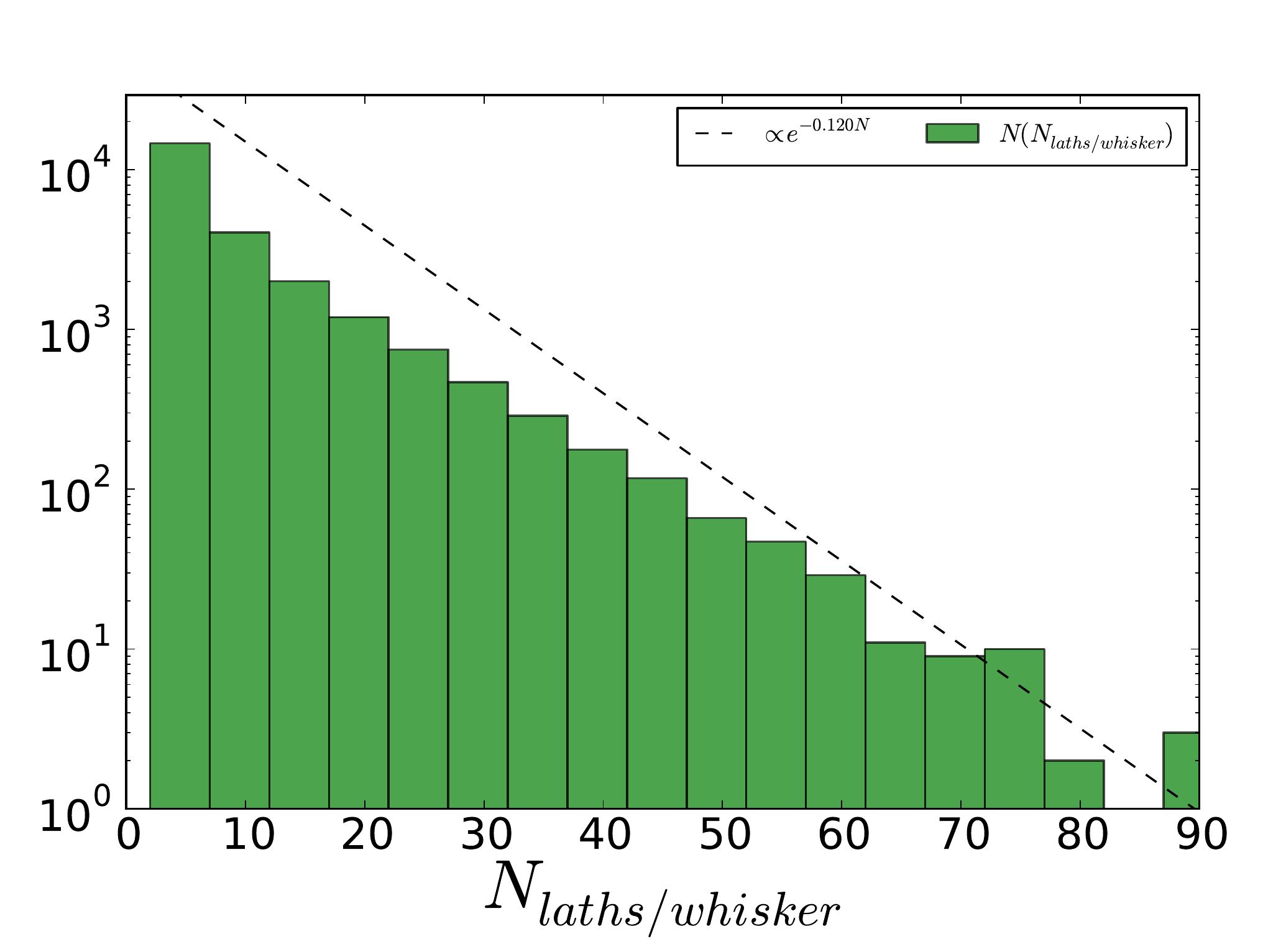} %FIG9
\caption{
\label{770_LOGhistoLATHS.pdf}
Histogram of laths per whisker; for $10$ different seeds of the random
number generator; previously equilibrated by $40$ $s$, and continued
by $40$ $s$; at a temperature of $80$ $K$; each simulation having
$512$ laths. The line is a function $\propto e^{-0.121 N}$, indicating
an exponential distribution with $\lambda = 1/\overline
N_{laths/whisker} = 1./8.28$, mean $\overline N_{laths/whisker} =
8.28$. The total number of sampled whiskers for this plot is $23835$.}
\end{figure}

Next, we are interested in the length-distributions of the whiskers in
equilibrium. In \ref{770_LOGhistoLATHS.pdf} we present the
distribution of lengths in boxes with temperature $T=80$ $K$. The plot
is based on the results of $10$ simulations, starting with different
seeds for the random number generators and equilibrated for $40$ $s$,
after which production runs of $40$ $s$ each were started. As is clear
from the plot, the lengths of whiskers are exponentially distributed,
with a mean value of $8.28$ laths per whisker. This results in a slope
in \ref{770_LOGhistoLATHS.pdf} of $\lambda = 1/8.28 \approx
0.121$. The exponential distribution allows for only very few long
whiskers; in the present case the longest whiskers found have lengths
of about $90$ laths. Nevertheless, such long whiskers quickly deplete
the pool of laths, which leads to severe finite size effects for
temperatures of about $80$ $K$ or less. This is clearly seen in
\ref{836__LOGlathsINwhiskerAFO_invTEMP.pdf} and \ref{836__RodsINwhiskerAFO_TEMP.pdf} where we have plotted the average lengths of
whiskers as a function of inverse temperature. The distribution
roughly behaves as described in the paper by Cates and Candau on
worm-like micelles \cite{0953-8984-2-33-001}. In this case the expressions relating the average number of laths per whisker and the temperature are:
\begin{eqnarray}
\nonumber  \ln\left(N_{laths/whisker} -2\right) = \frac{1}{2} \ln \phi +\frac{\epsilon}{2k_BT} \ \ \mbox{ for low temperatures }\\
  \ln\left(N_{laths/whisker} -2\right) =  \ln \phi +\frac{\epsilon}{k_BT} \ \ \mbox{ for high temperatures }
\end{eqnarray}

Our system recovers, quantitatively, the expected trend for low
temperatures.  The discrepancy with the theory may come from the fact
that the boxes have very low number of laths in the simulation box
compared to the real system; with only 512 of them, so there may be
size effects. This effect is also amplified if we take into the fact
that appearance of large whiskers is a rare event.

\begin{figure}[htbp]
\centering \includegraphics[width=14cm,bb=0 0 842 595]{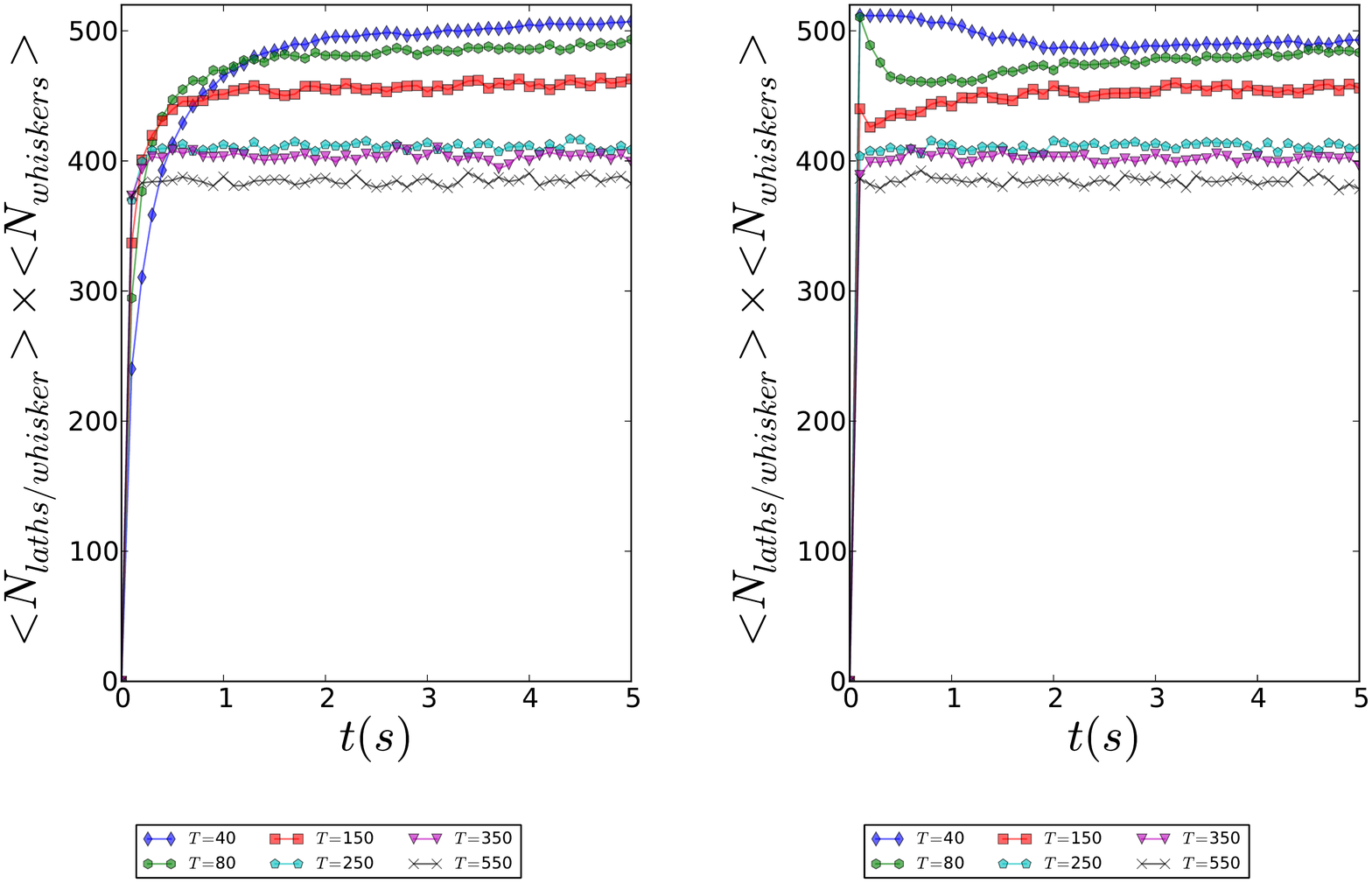} %FIG10
\caption{
\label{FIG10.pdf} 
Total material in whiskers (as the product of the average number of
whiskers and the average laths per whisker), for different
temperatures (in Kelvin); as a function of time and for the first $5$
$s$ of the evolution. \textit{(Left)} Initially disordered
configuration. \textit{(Right)} Initially ordered configuration.  (\textit{Color online.})}
\end{figure}

\begin{figure}[htbp]
\centering
\includegraphics[width=8cm,bb=0 0 585 441]{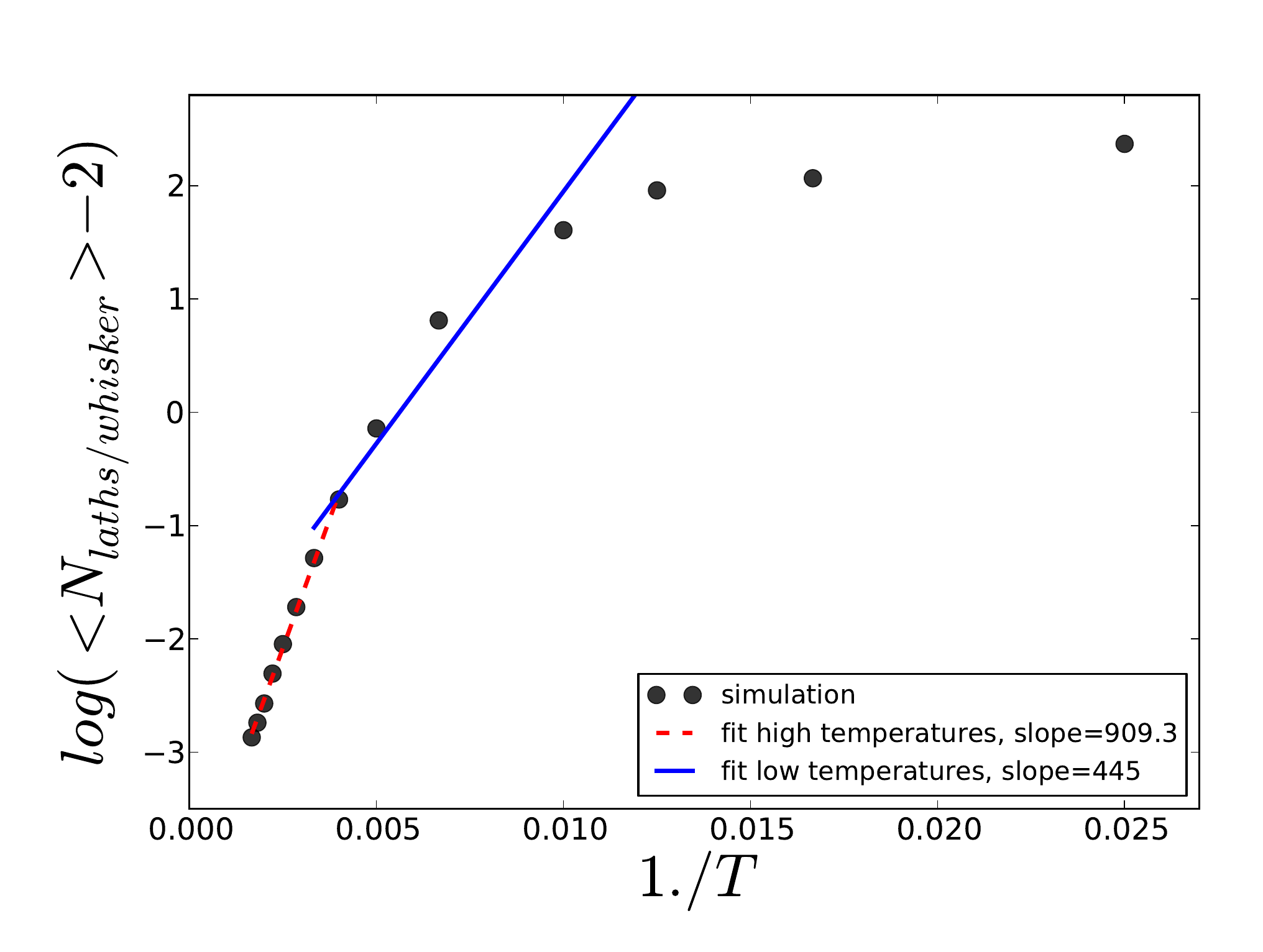} %FIG11
\caption{
\label{836__LOGlathsINwhiskerAFO_invTEMP.pdf}
Average number of laths per whisker, against $1/T$. The simulation was
run for $40 s$ for equilibration and then continued for another $100$
$s$. $T$ in Kelvin.  (\textit{Color online.})}
\end{figure}

\begin{figure}[htbp]
\centering
\includegraphics[width=8cm,bb=0 0 585 441]{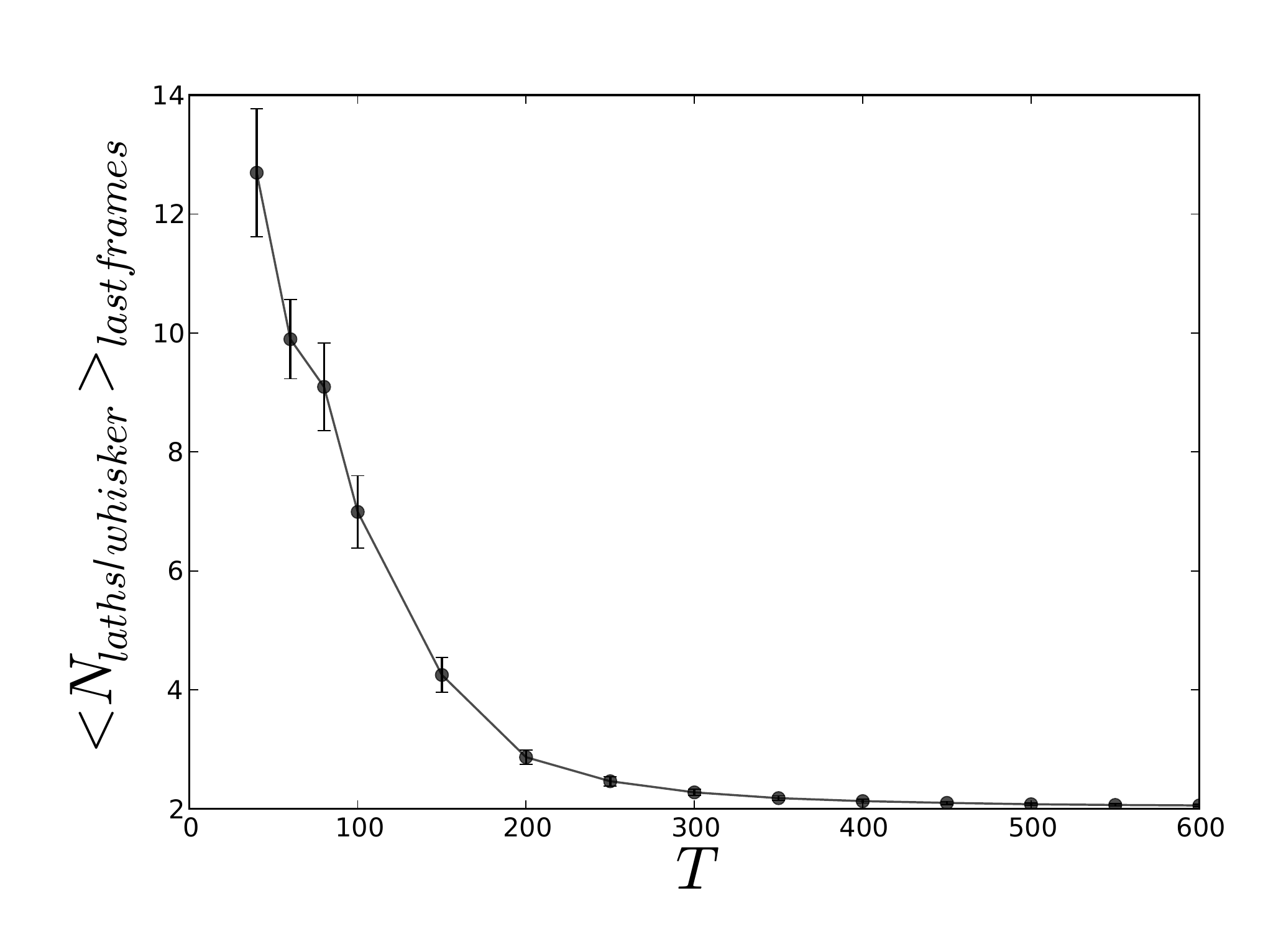} %FIG12
\caption{
\label{836__RodsINwhiskerAFO_TEMP.pdf}
Average number of laths per whisker. The simulation was run for $40 s$
for equilibration and then continued for another $100$ $s$. $T$ in
Kelvin. The independent variable is $T$.}
\end{figure}

\subsection{Rheology and gel transition}

\begin{figure}[htbp]
\centering
\includegraphics[width=8cm,bb=0 0 585 441]{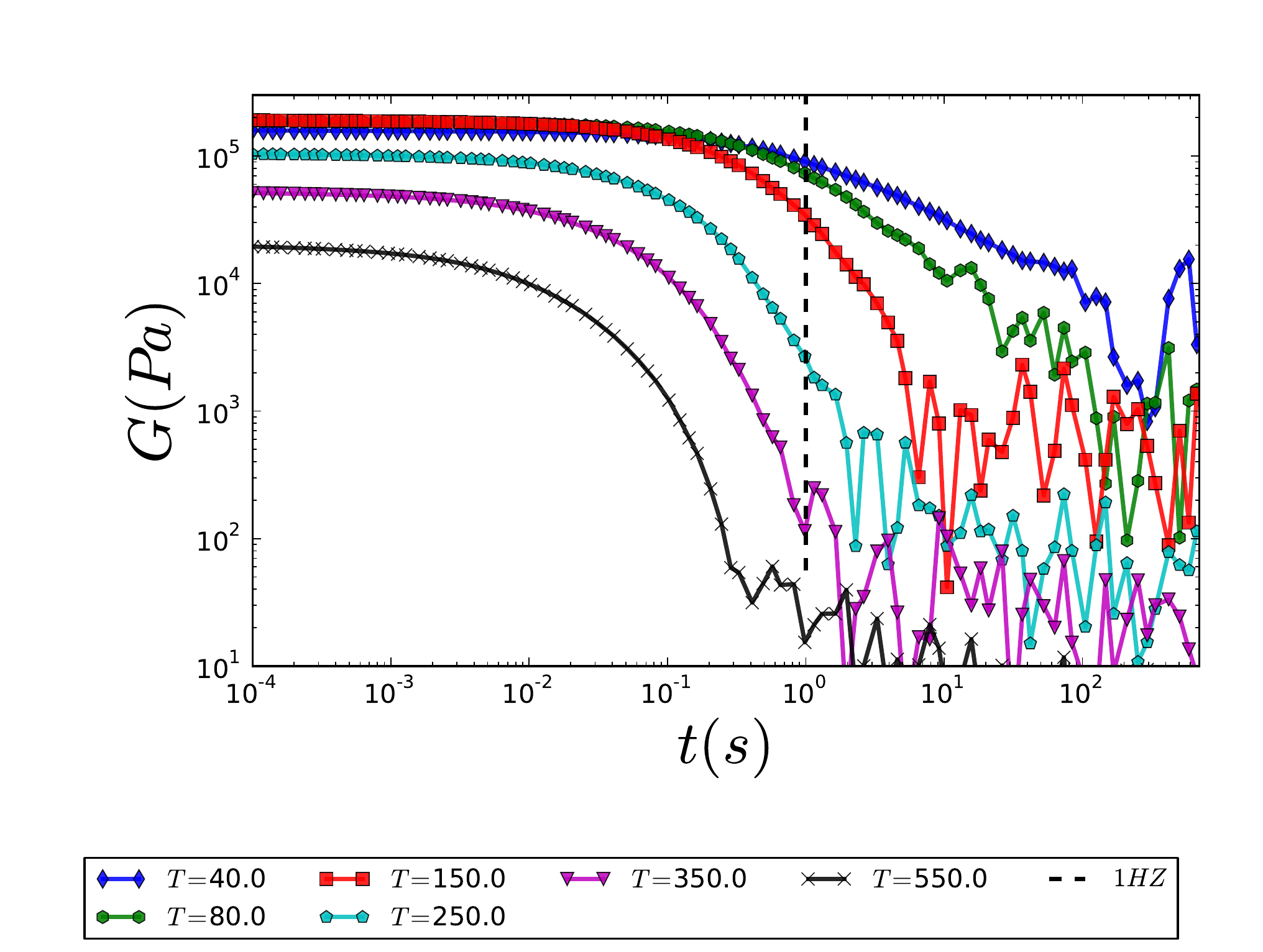} %FIG13
\caption{
\label{862_GofTIMEseveral.pdf}
Shear relaxation modulus,
  $G(t)$, for different values of the temperature of the system. The
  system equilibrated during (40 $s$); this data was taken after it
  had run for another 100 $s$. $T$ in Kelvin.  (\textit{Color online.})}
\end{figure}

In this section we study the rheological properties of our system. In
particular we will investigate how the shear relaxation modulus varies
with temperature, and how these variations are related to changes of
the distribution of whiskers.

In \ref{862_GofTIMEseveral.pdf} we present the shear relaxation
modulus $G(t)$ as a function of time for various temperatures. Since
we are not aiming to study rheological properties like the shear
storage and loss moduli, $G'$ and $G"$ respectively, in great detail,
we did not push calculations of the shear relaxation modulus to
convergence for times larger than a few seconds. It is clear from
\ref{862_GofTIMEseveral.pdf} that the plateau value at early times
quickly increases with decreasing temperature.
\begin{figure}[htbp]
\centering
\includegraphics[width=8cm,bb=0 0 585 441]{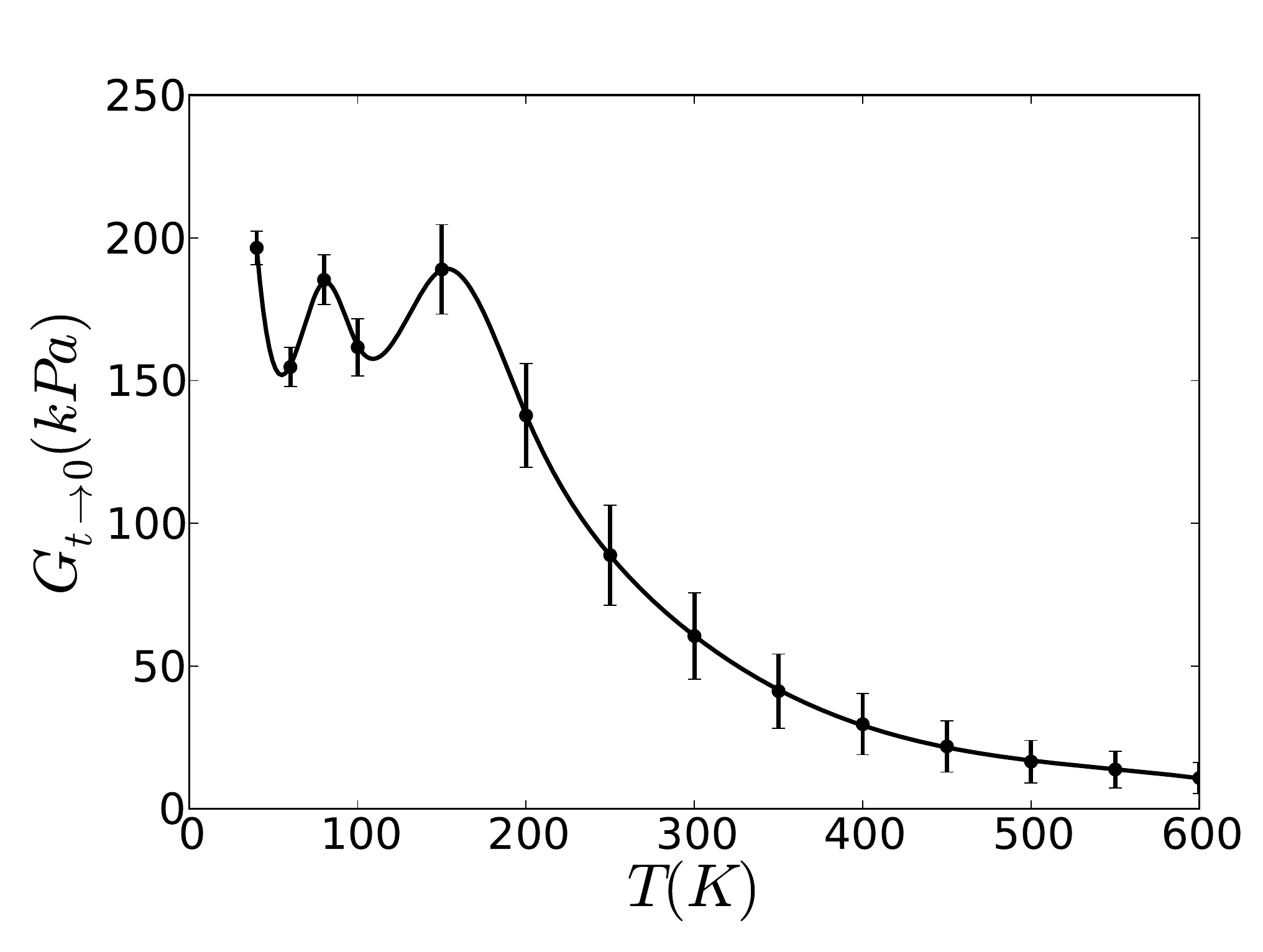} %FIG14
\caption{
\label{835_GofTEMPseveral.pdf}
Plateau value for $G$, as
  function of $T$.}
\end{figure}

\begin{figure}[htbp]
\centering
\includegraphics[width=7cm,bb=0 0 585 441]{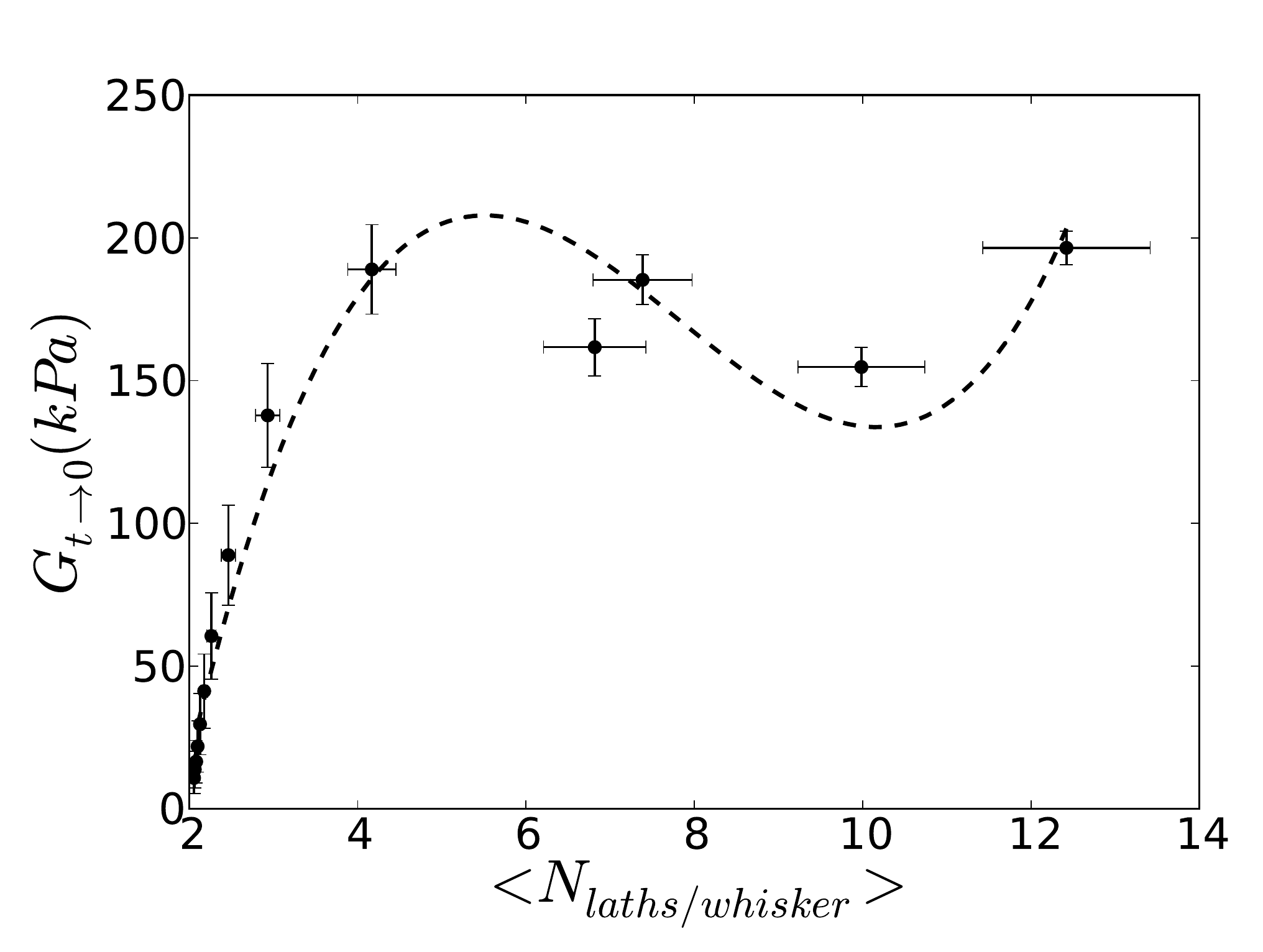} %FIG15
\caption{
\label{835_GofN_avseveral.pdf}
Plateau value for $G$, as
  function of the number length.}
\end{figure}

In \ref{835_GofTEMPseveral.pdf} we present these plateau values
as a function of temperature. As is clear from this plot, the zero
shear plateau values smoothly increase with decreasing time, up to
temperatures of about 200 K. At lower temperatures they fluctuate
around an average value of about $170$ kPa. This is probably due to
finite size effects as the exponential length distribution becomes
severely perturbed since the box can not accommodate sufficiently long
whiskers. This has already been noticed above when discussing the
average whisker length as function of temperature.

In \ref{835_GofN_avseveral.pdf} we plot the same data, but now
against the average number of laths per whisker. For small values of
$<N_{laths/whisker}>$, $G_{t\rightarrow 0}$ smoothly increases with
increasing $<N_{laths/whisker}>$. For large values of the latter, the
zero shear plateau fluctuates around a constant value.

\begin{figure}[htbp]
\centering
\includegraphics[width=8cm,bb=0 0 585 441]{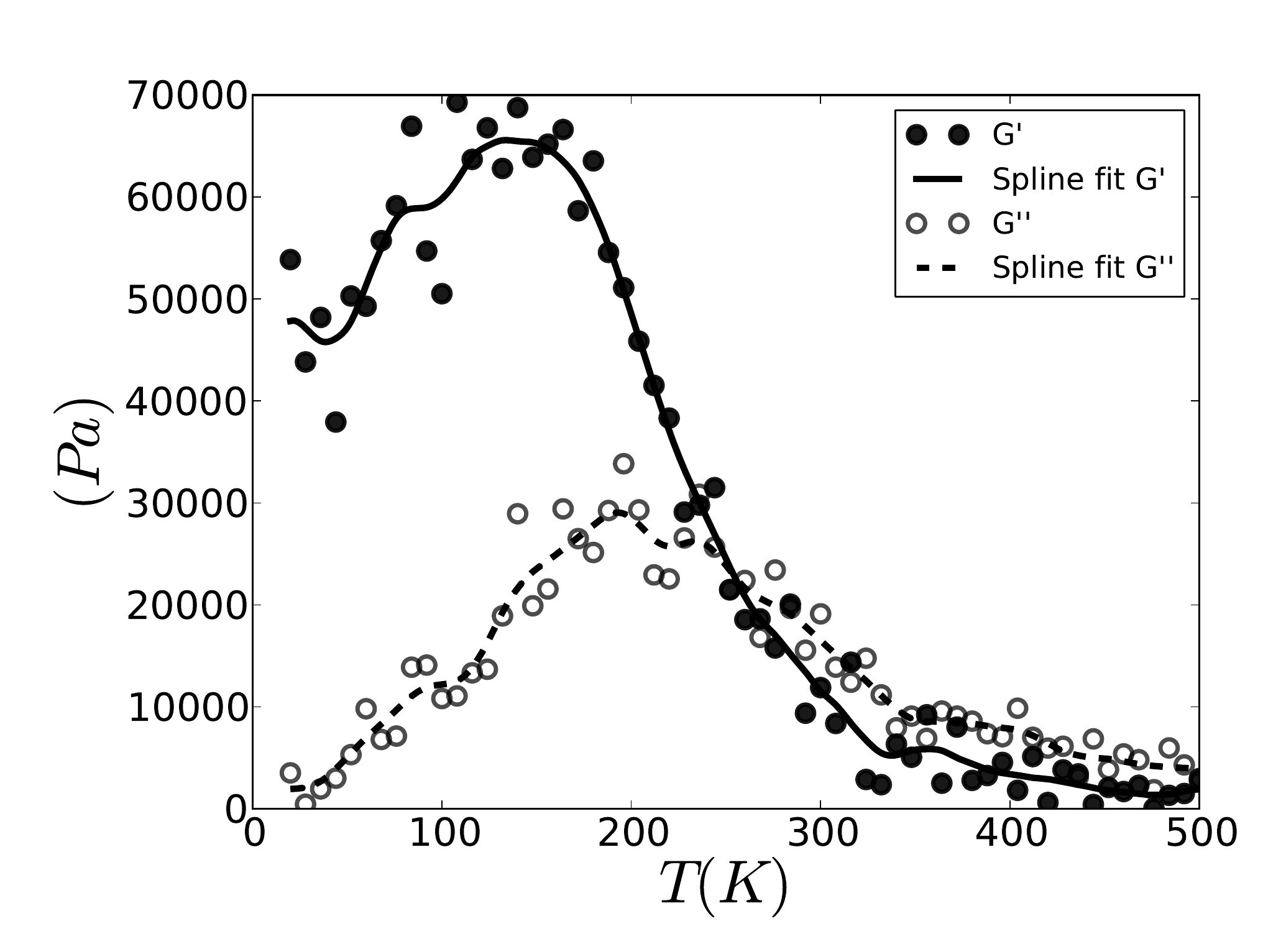} %FIG16
\caption{
\label{compare_791_with_812_fitGpGppAFOT.pdf}
Oscillatory
  shear numerical experiment. We ran for
  $60\times 10^5$ time steps (60 $s$) discretely increasing the
  temperature each $1\times 10^5$ steps (1 $s$), in 8 $K$ each
  time. This gives an average rate of 8 $K/s$.}
\end{figure}

Next we recall that the experimental determination of the gel
transition is usually based on values of $G'$ and $G''$ at a frequency
of one Hertz. If we now return to the shear relaxation modulus as a
function of time (see \ref{862_GofTIMEseveral.pdf}), we notice
that at times less than about one second high temperature $G(t)'s$
have decayed to very small values, while low temperature $G(t)'s$ have
not decayed at all. At intermediate temperatures the shear relaxation
modulus for $t=1 s$ sharply rises. Therefore, in
\ref{compare_791_with_812_fitGpGppAFOT.pdf} we plot the
storage and loss moduli at one Hertz as a function of temperature.  It
is clear from this plot that for temperatures below about $200 K$ the
system behaves basically elastic, while for temperatures above $300 K$
the system is predominantly viscous. We therefore conclude that
according to the usual rheological definition a gel transition occurs
at about $250 K$. Given the fact that our model has to be taken as highly coarse grained with relation to polymers, we find it remarkable the fact that this behavior is quite similar to that reported in rheological experiments of gelation of P3HT \cite{doi:10.1021/ma202564k}. The main difference of our model with that reference is the absence of hysteresis in our model. It may be caused by the lack of branching of our system, which can be studied in the future.

From the above results we come to the conclusion that the gel
transition in whisker forming systems is not a real transition, but
depends very much the definition being used. Moreover, being used to
different strengths of everyday forces and times will define the gel
transition using a different frequency, and arrive at much lower gel
transition temperatures.

\section{Summary}

In this letter we have presented a coarse grained model used to study
the whisker formation of long slender colloidal sized particles. The model is general, but we argue that it can be also seen as a highly coarse grained model for polymers. In that sense, it can be
applied to the aggregation happening due to the $\pi - \pi$ stacking
interaction between polythiophenes. For that system, in the crystallized
state of the molecule the planes defined by its backbone 
lie parallel to each other, forming long threads (up to the
$\mu m$ scale), called \emph{whiskers} in the literature.  For
example, a P3HT molecule of of contour length of 65 nm, produces
whiskers of $15 \sim 20$ nm width, 5nm thick -two or three chain
layers-, and $10 \mu m$ long \cite{Lim201014}. Here our 
coarse grained model is a tool to study the rheology, but disregarding
the internal flexibility of the molecule (specifically, we do not
simulate the hair-pin structure, in which the P3HT bends while
crystallizing reaching half of its length. This can be studied in a
follow up research).

We propose to model the molecule as a long slender plane, i.e., with a
high aspect ratio between the two in-plane directions; which we call a
\emph{lath}. A lath is defined to remain \emph{free} if the energy of
interaction with all its neighbors is above some minimum threshold
(small in absolute value or positive). On the contrary a lath within a
whisker is \emph{stuck} to its two closest neighbors and does not
interact with the others.

The potential energy depends on a single energetic parameter, and also
on the history of the system; having slightly different expressions
for two, three and four body interactions. This allows for a very fast
algorithm, that can be used to study the rheology of the system.
Given our choice of the parameters, we obtain a whisker formation
process that reaches an equilibrium state in matter of twenty to
twenty five seconds, depending on the temperature. The distribution
of number of laths per whisker, which depends on the temperature, has
an exponential shape. For $T=80$ $K$, this gives a mean of $8.28$ laths
per whisker; similar to the case of worm-like micelles.

Our system, while being simple, does reproduce the gel transition
that happens in the case of P3HT in solution. An oscillatory shear
experiment that measures $G'$ and $G''$ while slowly decreasing the
temperature $8$ $K$ each $1$ $s$, gives rise to a transition from
liquid to gel (imposing a decrease in temperature implies that it is
not an strictly closed system, so its entropy does not need to
increase). As the transition is the product of the formation of the
whiskers, we conclude that part of the rheological changes can be
accounted only to the fact of the whisker formation. The gelation
process is also seen experimentally, thus we are confident that our
model captures the basic physics of the system. 

It has been reported in the literature an hysteresis between the
ramping up and down of the gelation process
\cite{doi:10.1021/ma202564k}. Our system does not show this
behavior. A possible reason is the absence of branching in our
system. Not having this possibility makes the system pseudo-one
dimensional; and since there are no phase transitions for one
dimensional structures, while for higher dimensional there are both
percolation and phase transitions, it is expected that branching would
provide such transitions.

Further improvements to the present model include the study of the
interaction between sphere-like particles, like the PCBM or C60, with
the laths. It could also be important to study the effect on the
alignment that can be obtained by including a second kind of lath,
that not necessarily shares the aromatic interaction; as for instance
 gold nanorods \cite{doi:10.1021/ma402179w}.

\section{Conclusion}

We have investigated the gel transition in $\pi$-stacking long laths,
being prototype for moderately long stiff aromatic polymers. The
$\pi$-stacking bonds allow for the creation of long whiskers of
consecutively stacked laths.  We did not allow for branching of such
whiskers, thereby preventing the occurrence of extended
three-dimensional networks.  In particular, the pseudo one-dimensional
character of the whiskers does not allow for structural transitions
with varying temperatures that could depend on the direction of the
temperature gradient; in other words, hysteresis is not present in the
gel-sol transition. This character also forces their length
distribution to be exponential.  As a result, only very few long
whiskers occur amidst a see of many smaller whiskers.  Chances of
having mechanically percolating structures are therefore very low.
Nevertheless, we have found that with the usual definition of gel
transition, occurring at a temperature where the ratio of the storage
and loss moduli at a frequency of one Hertz changes from values larger
than one to values lower than one, such a transition indeed does
occur. Our main conclusion therefore is that the occurrence of a gel
transition does not necessarily imply that percolating three
dimensional structures have developed in the system.

\section{Acknowledgments}

Some of the simulations for this paper used computational time from
the ``Centro de Informatica y Biolog\'\i a Computacional BIOS'', and
ran through the ``Red Nacional Acad\'emica de Tecnolog\'\i a Avanzada,
RENATA''. This work was partially funded by Ozofab, EOST-10023-OZOFAB;
and ESMI, Grant Agreement No. 262348 ``ESMI''. I want to thank current
and former members of the Computational Biophysics Group, specially to
Professor Wim Briels, whose dedication, insight and hard work improved
the quality of this paper; Professor Wouter den Otter for deep,
imaginative and very fruitful discussions; and to Igor Santos de
Oliveira, for sharing his knowledge on computational rheology.

%\appendix{Forces and Torques}
%\section{Forces and Torques}
%\subsubsection{Forces}

\section{Bibliography}
\bibliographystyle{plain}
\bibliography{boldarticle}
\printindex

\section{Appendix. Forces and Torques}
\subsection{Forces}
%\appendix
The force acting on lath $i$ is: 
\bea \nonumber
\vect F = -\derp{\Phi_S}{\vect R_i}. 
\eea

As the potential is the product of different functions is useful to go
over their derivatives with respect to $R_i$. Take first $V_{d}$: 
\bea
\nonumber -\derp{V_{d}(r_{kj})}{\vect R_i} = \frac{\epsilon}{2} \left(
\frac{-a}{\cosh^2(a(r_{kj} - \frac{\sigma}{2}))\tanh \left(\frac{a
    \sigma}{2}\right)} \right. \\
%\frac{1}{}
\left. \times \frac{\vect r_{jk}}{|\vect r_{jk}|}(\delta_{ik} - \delta_{ij}) \right).
\eea

The $V_{cn}$ potential takes two forms, depending on the $\uni n_k$
vector, either : 
\beq \label{eq:Vso2Body}(\uni n_{k,j,k} \cdot \uni
r_{k,j})^{2l},\eeq 
\noindent or \beq \label{eq:Vso3Body} (\uni n_{\neig(k),k,k} \cdot \uni
r_{k,j})^{2l}.\eeq

One way to calculate the stress components during the simulation,
taking into account the periodic boundary conditions, requires to
explicitly write the forces as derivatives with respect to the
relative vectors $\vect r_{kj}$ and $\vect r_{\neig(k),k}$. As
\eqref{eq:Vso2Body} depends on $\vect r_{kj}$ (there is no $\neig(k)$
lath to take into account here), then the derivative with respect
to either $k$ or $j$ is: 
%\begin{strip}
\bea \nonumber -\derp{}{\vect R_i} \left(
\frac{\vect n_{k,j,k}\cdot \vect
  r_{k,j}}{|\vect n_{k,j,k}||\vect r_{k,j}|}\right)^{2l} &=& \left[ \frac{4l
    (\vect n_{k,j,k}\cdot \vect r_{k,j})^{2l -1
  }}{(|\vect n_{k,j,k}||\vect r_{k,j}|)^{2l}} \vect n_{k,j,k}\right.\\ &-&  
  \left. \frac{2l(\vect n_{k,j,k} \cdot \vect
    r_{k,j})^{2l}}{(|\vect  n_{k,j,k}||\vect  r_{k,j}|)^{2l+1}}\left(|\vect  r_{k,j}| \uni
  n_{k,j,k} + |\vect  n_{k,j,k}| \uni r_{k,j}\right)\right](\delta_{i,k} -
\delta_{i,j}). 
\eea

\noindent In the case that the potential is \eqref{eq:Vso3Body},
the derivative with respect to the coordinates of lath $k$ is:
\bea \derp{}{\vect R_i}
\left(\frac{\vect n_{\neig(k),k,k} \cdot \vect
  r_{k,j}}{|\vect n_{k,j,k}||\vect r_{k,j}|}\right)^{2l} = \derp{}{\vect r_{kj}}
\left(\frac{\vect n_{\neig(k),k,k} \cdot \vect
  r_{k,j}}{|\vect n_{k,j,k}||\vect r_{k,j}|}\right)^{2l} \derp{\vect r_{kj}}{\vect
  R_k} + \derp{}{\vect r_{k}} \left(\frac{\vect n_{\neig(k),k,k} \cdot
  \vect r_{k,j}}{|\vect n_{k,j,k}||\vect r_{k,j}|}\right)^{2l} \derp{\vect
  r_{\neig(k),k}}{\vect R_k}, \eea

\noindent Obtaining:
\bea
\derp{}{\vect r_{kj}}
\left(\frac{\vect n_{\neig(k),k,k} \cdot \vect
  r_{k,j}}{|n_{k,j,k}||r_{k,j}|}\right)^{2l}  &=& \nonumber \frac { 2l
  [ \nskk \cdot \vrkj]^{2l-1}}{[|\vrkj||\nskk|]^{2l}} 
\nskk -
\frac {2l \starx^{2l}}{[|\vrkj||\nskk|]^{2l+1}}
  |\nskk| \uni r_{k,j} 
\\ \\
\derp{}{\vect r_{\neig(k),k}}
\left(\frac{\vect n_{\neig(k),k,k} \cdot \vect
  r_{k,j}}{|n_{k,j,k}||r_{k,j}|}\right)^{2l} &=&
\nonumber \frac { 2l [ \nskk \cdot \vrkj]^{2l-1}}{[|\vrkj||\nskk|]^{2l}} \vect n_{k,j,k} - \frac {2l \starx^{2l}}{[|\vrkj||\nskk|]^{2l+1}}
|\vrkj| \uni n_{\neig(k),k,k},\\
\eea

\noindent and :$\derp{\vect r_{\neig(k),k}}{\vect R_k} = 1$,
$\derp{\vect r_{kj}}{\vect R_k} =-1$. The derivative with respect to
$j$,$k$ or $\neig(k)$ is:
\bea \nonumber \nonumber -\derp{}{\vect
  R_i} \left(\frac{\vect n_{\neig(k),k,k} \cdot \vect
  r_{k,j}}{|n_{k,j,k}||r_{k,j}|}\right)^{2l} &=& \nonumber \frac { 2l
  [ \nskk \cdot \vrkj]^{2l-1}}{[|\vrkj||\nskk|]^{2l}} \left( \vect
n_{j,k,k} (\delta_{i,k}-\delta_{i,\neig(k)}) + \nskk (\delta_{i,k} -
\delta_{i,j}) \right)\\ \nonumber&+& \frac {2l
  \starx^{2l}}{[|\vrkj||\nskk|]^{2l+1}}\left(|\vrkj| \uni
n_{\neig(k),k,k}(\delta_{i,k} - \delta_{i,\neig(k)}) - |\nskk| \uni
r_{k,j} (\delta_{i,k} - \delta_{i,j} ) \right), \\ \eea
%\end{strip}

\subsection{Torques}
\appendix
The definition of the torque is:
\bea \nonumber \mathbf T_{i,stk.} &=& -\orii \times \derp{\Phi_S}{\orii}. \eea

\noindent And given the torque, the evolution of the orientation
vector is given by: \beq d\ori = \underbrace{[\hat I - \ori \ori]
  \cdot \Gamma \cdot \ori dt}_{\delta u_{i_1}} +
\underbrace{\frac{1}{\gamma} \vect T \times \ori dt}_{\delta u_{i_2}}
\eeq

\noindent With $\delta_{u_1}$ related to the flow. Both $V_{o}$ and
$V_{cn}$ depend on the orientation $\uni u$. 

%\begin{strip}
\bea
- \left( \orii \times \derp{\vso}{\orii} \right) \times \orii 
\nonumber &=& 2l\delta_{i,k} \left(\frac{\starx^{2l-1}}{[|\nskk||\vrkj|]^{2l}}\left(
\nskk (\vrkj \cdot \orik) + \nkj(\orik \cdot \vrnkk)\right)\right.\\ &-& \left.
\frac{ \starx ^{2l}}{|\nskk|^{2l+2}|\vrkj|^{2l}}\nskk (\vrnkk \cdot \orik)  \right)
\label{eq:3btorqueSov1}
\eea

For the case in which the orientation vector $\vect n_k$ is not given
by a stuck lath the potential is $V_{cn}(\uni n_{k,j,k}\cdot \uni
r_{k,j})$, and the analogous calculation gives:
\bea
&-& \left( \orii \times \derp{\vso}{\orii} \right) \times \orii =\\
\nonumber &2l& \delta_{i,k} \left(\frac{(\vect n_{k,j,k}\cdot \vect r_{k,j})^{2l-1}}{[|\vect n_{k,j,k}||\vrkj|]^{2l}}
2 \vect n_{k,j,k} (\vrkj \cdot \orik) - %\right.\\ &-& \left.
\frac{ (\vect n_{k,j,k} \cdot \vect r_{k,j}) ^{2l}}{|\vect n_{k,j,k}|^{2l+2}|\vrkj|^{2l}}\vect n_{k,j,k} (\vect r_{k,j} \cdot \orik)  \right)
\label{eq:3btorqueSov2}
\eea
%\end{strip}

%\newpage
%\begin{figure}[htbp]
%\centering
%\includegraphics[width=8.3cm,bb=0 0 1694 714]{table_of_contents_graphic.jpg} %table of contents graphic
%\captionsetup{labelformat=empty}
%\caption{
%\label{table_of_contents_graphic.jpg}
%For Table of Contents use only
%}
%\end{figure}

\end{document}